\documentclass{optica-article}

\journal{opticajournal} 

\articletype{Research Article}

\usepackage{lineno}

\usepackage{amsmath,amsfonts,bm}

\newcommand{\imag}{\mathrm{i}}

\begin{document}

\title{On the modeling and mitigation of interference fringes in polarimetric instrumentation}

\author{R.\ Casini\authormark{1,*} and D.~M.\ Harrington\authormark{2}}

\address{\authormark{1}NSF NCAR-HAO, 3090 Center Green Dr., Boulder, CO 80301, USA\\
\authormark{2}NSO-DKIST, 22 Ohia Ku St., Pukalani, HI 96768, USA}

\email{\authormark{*}casini@ucar.edu} 


\begin{abstract*}
Spectral and spatial fringes in polarized light are produced by the interference of transmitted and reflected waves at the interface between materials with different indexes of refraction. These instrumental artifacts can affect the accuracy of optical designs conceived for high-sensitivity spectroscopy and polarimetry. We consider the principal sources of these artifacts and the possible design pathways to mitigate them. In order to do so, we have developed an approximate yet agile treatment of the problem of the transmission and reflection of light in birefringent materials, which fundamentally relies on the assumption of \emph{small birefringence} of the modeled materials for its implementation. The comparison of our results with those from more rigorous treatments, such as Berreman calculus, thus also serves as a test of the limits of the small-birefringence approximation in optical design applications. 
The treatment presented in this work is limited to isotropic materials and uniaxial crystals, which are the most common types of optics employed in polarimetric instrumentation.
An extensive set of modeling examples is provided to illustrate the salient characteristics of polarization fringes and their dependence on optical design parameters.
\end{abstract*}

\section{Introduction}

Spectro-polarimetry, i.e., the spectral analysis of the polarization signatures in the radiation we receive from an observed target, is the primary diagnostic tool to investigate the physical causes of symmetry breaking in the process of light emission from objects of scientific interest, causing the emitted radiation to be polarized. Often these signals have very weak amplitudes (e.g., well below 1\% of the radiation intensity), requiring spectro-polarimetric instrumentation with sufficient precision and accuracy to correctly discern and interpret the measured signals.

Among the principal causes of systematic errors in spectro-polarimetry is the formation of both spectral and spatial fringes by the interference of transmitted and reflected light waves through the optical system, the amplitude and frequency of which can occur at intensity and frequency scales that directly impact the accurate recording and successive interpretation of the polarized spectral signals that constitute the main diagnostics of the observations. The ability to model the formation of such interference effects in the recorded signals, at least to the extent that their impact can be estimated with sufficient precision to identify alternative or corrective instrument design choices, is paramount for the success of observing programs that rely on such techniques. The NSF Daniel K.\ Inouye Solar Telescope \cite{Ri22} includes a suite of state-of-the-art spectro-polarimeters for the investigation of solar magnetism, the design of which have benefited from the ability to realistically model the behavior of polarization optics such as the production of polarization fringes \cite{JATIS6,Ha23}; these modeling efforts have relied on the rigorous treatment of the problem originally formulated by \cite{Be72} (see also \cite{MC15}). These exact albeit much more laborious works remain the standard reference for modeling general problems of birefringent optics.

In this paper we propose an approximate treatment of this problem, which allows us to rapidly estimate interference effects in optical systems consisting of isotropic materials or uniaxial crystals, with a verified degree of accuracy that is sufficient to inform the design of spectro-polarimeters, aiming at the mitigation of such effects. We rely on existing literature on the subject to base the formal derivation of our modeling approach, and we present various application examples to clarify the production mechanisms and main drivers of polarization fringes.

\begin{figure}[t!]
\centering
\includegraphics[width=.8\textwidth]{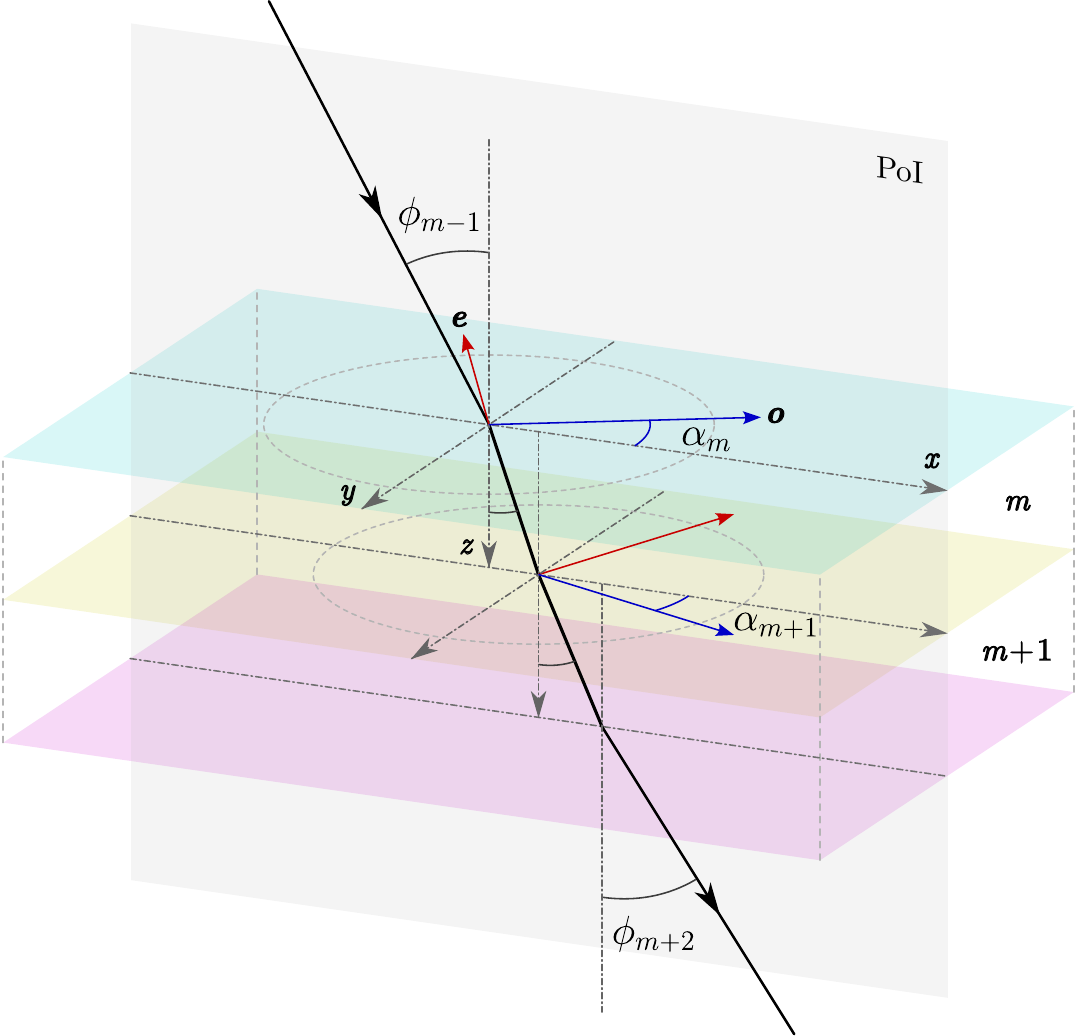}
\caption{\label{fig:perp_geom}
Propagation of a light ray through a stack of plane-parallel uniaxial crystals, rotated at different angles $\alpha$ with respect to the plane of incidence (PoI) $\langle x,z\rangle$ (in this example, $\alpha_m$ is negative, while $\alpha_{m+1}$ is positive). The $xyz$ axes provide the natural reference for the polarization decomposition of the incoming beam, with $y\parallel\bm{s}$, and $\bm{p}\in\langle x,z\rangle$. The figure shows the case of a stack of uniaxial crystals that have all been cut such that the optic axes ($\bm{e}$) are parallel to the interfaces. This is a common requirement for birefringent optics employed in polarimetric instrumentation, as it maximizes the birefringence experienced by the beam while minimizing polarization aberrations due to the phenomenon of double refraction.
For a \emph{positive} uniaxial crystal, the $\bm{o}$ and $\bm{e}$ axes in the figure correspond, respectively, to the fast and slow axes of the crystal, and they also correspond to orthogonal states of linear polarization for a ray propagating along $z$;
for a \emph{negative} uniaxial crystal, $\bm{o}$ becomes the slow axis and $\bm{e}$ the fast axis. The figure 
also clarifies the convention followed in this work, where the angle $\alpha$ of the optic is defined by the orientation of the $\bm{o}$ axis.
In our treatment, we will always assume the approximation of small birefringence \cite{Ye82}, 
so the ordinary (o) and extraordinary (e) rays, into which an incoming ray splits at the interface with a birefringent medium, are assumed to always lie on the PoI, and with polarizations that are orthogonal to each other, \emph{regardless of the angle $\phi$ of incidence  and the orientation $\alpha$ of the optic.}
}
\end{figure}

\section{Jones calculus approach to polarization fringes} \label{sec:formalism}
We treat the problem of modeling the formation of polarized fringes via
transmission and reflection in a stack of plane-parallel layers starting from the treatment of the case of isotropic materials as described by \cite{He91}, \S4.8.\footnote{It is worth noting that \cite{He91} provides a non-conventional sign choice for the Fresnel coefficient $r^p$, compared to, e.g., the monograph by \cite{Di63}, which also treats this problem extensively in Chaps.~14-15.}.
%
In particular, the matrix expression provided by Eq.~4(100) in that reference serves as the starting point of the present development. Accordingly, we follow as closely as possible the same notation of \cite{He91}, and write the transfer operation of the radiation electric vector across the interface between two ordered elements, $(m-1)$ and $m$, as
\begin{equation} \label{eq:tranmat}
\begin{pmatrix}
E_{m-1}^+ \\ E_{m-1}^-
\end{pmatrix}
=\frac{1}{t_m}
\begin{pmatrix}
{\rm e}^{\imag \delta_{m-1}}
&r_m\,{\rm e}^{\imag \delta_{m-1}} \\
r_m\,{\rm e}^{-\imag \delta_{m-1}} 
&{\rm e}^{-\imag \delta_{m-1}}
\end{pmatrix}
\begin{pmatrix}
E_m^+ \\ E_m^-
\end{pmatrix}
\equiv \mathbf{M}_{m-1,m}
\begin{pmatrix}
E_m^+ \\ E_m^-
\end{pmatrix}\;,
\end{equation}
where with $+$ and $-$ we indicate the transmitted and reflected components of
the electric field, respectively,
and $t_m$ and $r_m$ are the Fresnel
coefficients for transmission and reflection at the interface between
the two elements. These coefficients are given by
expressions such as Eqs.~4(19--22) of \cite{He91}, along with their
generalization to complex refractive indexes $\nu_m=n_m-\imag k_m$ for absorptive materials,
propagation angles $\phi_m\ne 0$ with respect to the surface's normal, 
and for the possible existence of different refractive indexes in the case of
anisotropic media (here limited to uniaxial crystals). We provide a compendium of the relevant formulas in App.~\ref{app:numerical}.

In Eq.~(\ref{eq:tranmat}), 
$\delta_m$ is the (generally complex) phase delay produced by the layer of index $\nu_m$ 
and thickness $d_m$, given by (cf.~\cite{He91}, Eq.~4(41))
\begin{equation} \label{eq:delta}
\delta_m=2\pi\,\frac{d_m}{\lambda}\,\nu_m\cos\phi_m\;.
\end{equation}
Equations~(\ref{eq:tranmat}) and (\ref{eq:delta}) are separately
specified for the $p$ and $s$ component of the field, with respect to
the reference frame of the principal plane of incidence (PoI), containing the
propagation direction $\bm{k}$ and the normal $\bm{n}$ to the (planar)
interface between the $(m-1)^\textrm{th}$ and $m^\textrm{th}$ elements. This is illustrated in Fig.~\ref{fig:perp_geom}, where the caption also provides a short summary of the various approximations involved.

Following \cite{Cl04c}, it is convenient to combine the $p$ and $s$
equations by introducing a $4{\times}4$ matrix formalism,
\begin{eqnarray} \label{eq:tranmat4}
\begin{pmatrix}
E_p^+ \\ E_p^- \\
E_s^+ \\ E_s^- 
\end{pmatrix}_{m-1}
&=&
\begin{pmatrix}
\mathbf{M}_{m-1,m}^p &\mathbf{0} \\
\mathbf{0} &\mathbf{M}_{m-1,m}^s
\end{pmatrix}
\begin{pmatrix}
E_p^+ \\ E_p^- \\
E_s^+ \\ E_s^- 
\end{pmatrix}_m \nonumber \\
&\equiv&\begin{pmatrix}
a_{m-1,m}^p &b_{m-1,m}^p &0 &0 \\
c_{m-1,m}^p &d_{m-1,m}^p &0 &0 \\
0 &0 &a_{m-1,m}^s &b_{m-1,m}^s \\
0 &0 &c_{m-1,m}^s &d_{m-1,m}^s 
\end{pmatrix}
\begin{pmatrix}
E_p^+ \\ E_p^- \\
E_s^+ \\ E_s^- 
\end{pmatrix}_m\;,
\end{eqnarray}
where the expressions of the coefficients $a$, $b$, $c$, and $d$ are
easily derived by comparison with Eq.~(\ref{eq:tranmat}).
Equation~(\ref{eq:tranmat4}) can be manipulated via row and column
permutations to give
\begin{equation}
\begin{pmatrix}
E_p^+ \\ E_s^+ \\
E_p^- \\ E_s^- 
\end{pmatrix}_{m-1}
=\begin{pmatrix}
a_{m-1,m}^p &0 &b_{m-1,m}^p &0 \\
0 &a_{m-1,m}^s &0 &b_{m-1,m}^s \\
c_{m-1,m}^p &0 &d_{m-1,m}^p &0 \\
0 &c_{m-1,m}^s &0 &d_{m-1,m}^s 
\end{pmatrix}
\begin{pmatrix}
E_p^+ \\ E_s^+ \\
E_p^- \\ E_s^- 
\end{pmatrix}_m\;,
\end{equation}
which can therefore be expressed in terms of the Jones vectors for the
transmitted and reflected beams, $\bm{J}^\pm\equiv(E_p^\pm,E_s^\pm)^T$,
where $(\cdot)^T$ is the operation of matrix transposition:
\begin{equation} \label{eq:jonestran4}
\begin{pmatrix}
\bm{J}^+ \\
\bm{J}^-
\end{pmatrix}_{m-1}
=\begin{pmatrix}
a_{m-1,m}^p &0 &b_{m-1,m}^p &0 \\
0 &a_{m-1,m}^s &0 &b_{m-1,m}^s \\
c_{m-1,m}^p &0 &d_{m-1,m}^p &0 \\
0 &c_{m-1,m}^s &0 &d_{m-1,m}^s 
\end{pmatrix}
\begin{pmatrix}
\bm{J}^+ \\
\bm{J}^-
\end{pmatrix}_m\;.
\end{equation}
Evidently, the
reference frame in which these Jones vectors are specified is the same
as described above, and is defined by the PoI and
the interface between the two materials $(m-1)$ and $m$.

One difficulty of the formalism is the fact that the matrices $\mathbf{M}_{m-1,m}$, or the coefficients $a_{m-1,m},\ldots,d_{m-1,m}$, mix 
the optical properties of the contiguous $(m-1)^\textrm{th}$ and $m^\textrm{th}$ materials. However, this situation 
can be somewhat ameliorated. First we write explicitly the transmission and reflection coefficients, 
\begin{align}
a_{m-1,m}^{p,s}&=\frac{{\rm e}^{\imag \delta_{m-1}^{p,s}}}{t_{m-1,m}^{p,s}}\;,
&d_{m-1,m}^{p,s}=(a_{m-1,m}^{p,s})^\ast\;, \\
\noalign{\vspace{3pt}}
b_{m-1,m}^{p,s}&=r_{m-1,m}^{p,s}\,a_{m-1,m}^{p,s}\;,
&c_{m-1,m}^{p,s}=(b_{m-1,m}^{p,s})^\ast\;,
\end{align}
where we indicated with $(\cdot)^\ast$ the operation of complex conjugation. Next, we observe that
%
\begin{equation} \label{eq:explmat}
\begin{pmatrix}
a_{m-1,m}^p &0 &b_{m-1,m}^p &0 \\
0 &a_{m-1,m}^s &0 &b_{m-1,m}^s \\
c_{m-1,m}^p &0 &d_{m-1,m}^p &0 \\
0 &c_{m-1,m}^s &0 &d_{m-1,m}^s 
\end{pmatrix}
\equiv \mathbf{\Lambda}_{m-1}^\ast\mathbf{O}_{m-1,m}\;,
\end{equation}
having defined
\begin{equation} \label{eq:Lmat-Omat}
\mathbf{\Lambda}_m\equiv
\begin{pmatrix}
{\rm e}^{-\imag \delta_m^p} &0 &0 &0 \\
0 &{\rm e}^{-\imag \delta_m^s} &0 &0 \\
0 &0 &{\rm e}^{\imag \delta_m^p} &0 \\
0 &0 &0 &{\rm e}^{\imag \delta_m^s}
\end{pmatrix}\;, \quad
\mathbf{O}_{m-1,m}\equiv\begin{pmatrix}
\frac{1}{t_{m-1,m}^p} &0 &\frac{r_{m-1,m}^p}{t_{m-1,m}^p} &0 \\
0 &\frac{1}{t_{m-1,m}^s} &0 &\frac{r_{m-1,m}^s}{t_{m-1,m}^s} \\
\frac{r_{m-1,m}^p}{t_{m-1,m}^p} &0 &\frac{1}{t_{m-1,m}^p} &0 \\
0 &\frac{r_{m-1,m}^s}{t_{m-1,m}^s} &0 &\frac{1}{t_{m-1,m}^s}
\end{pmatrix}\;,
\end{equation}
where now $\mathbf{\Lambda}_m$ only contains optical
properties pertaining to the $m$ element. On the other hand,
$\mathbf{O}_{m-1,m}$ still depends on both materials on the two sides 
of the interface, because of the structure of the Fresnel coefficients.
We note that $\mathbf{O}_{m-1,m}$ can further be split into the product 
of a ``reflection'' matrix with the inverse of a 
diagonal ``transmission'' matrix, being
\begin{eqnarray} \label{eq:Osplit}
\mathbf{O}_{m-1,m}
&=&
\begin{pmatrix}
1 &0 &r_{m-1,m}^p &0 \\
0 &1 &0 &r_{m-1,m}^s \\
r_{m-1,m}^p &0 &1 &0 \\
0 &r_{m-1,m}^s &0 &1
\end{pmatrix}
\begin{pmatrix}
t_{m-1,m}^p &0 &0 &0 \\
0 &t_{m-1,m}^s &0 &0 \\
0 &0 &t_{m-1,m}^p &0 \\
0 &0 &0 &t_{m-1,m}^s
\end{pmatrix}^{\!-1} \nonumber \\
&\equiv& \mathbf{R}_{m-1,m}\mathbf{T}_{m-1,m}^{-1}\;.
\end{eqnarray}

With these definitions, from Eq.~(\ref{eq:jonestran4}) we can write
\begin{equation} \label{eq:temp}
\mathbf{\Lambda}_{m-1}
\begin{pmatrix}
\bm{J}^+ \\
\bm{J}^-
\end{pmatrix}_{m-1}
=\mathbf{O}_{m-1,m}
\begin{pmatrix}
\bm{J}^+ \\
\bm{J}^-
\end{pmatrix}_m 
=\left(\mathbf{O}_{m-1,m}\, \mathbf{\Lambda}_m^\ast\right) 
\mathbf{\Lambda}_m
\begin{pmatrix}
\bm{J}^+ \\
\bm{J}^-
\end{pmatrix}_m\;,
\end{equation}
and by defining the \emph{phase-delayed} Jones 4-vector
\begin{equation} \label{eq:phase-jones}
\begin{pmatrix}
\bm{\hat J}^+ \\
\bm{\hat J}^-
\end{pmatrix}_m\equiv \mathbf{\Lambda}_m
\begin{pmatrix}
\bm{J}^+ \\
\bm{J}^-
\end{pmatrix}_m\;,
\end{equation}
we finally arrive at the expression
\begin{equation} \label{eq:jonestran4.alt}
\begin{pmatrix}
\bm{\hat J}^+ \\
\bm{\hat J}^-
\end{pmatrix}_{m-1}
=\mathbf{O}_{m-1,m}\,
\mathbf{\Lambda}_m^\ast 
\begin{pmatrix}
\bm{\hat J}^+ \\
\bm{\hat J}^-
\end{pmatrix}_m
=\mathbf{R}_{m-1,m}\mathbf{T}_{m-1,m}^{-1}\,
\mathbf{\Lambda}_m^\ast 
\begin{pmatrix}
\bm{\hat J}^+ \\
\bm{\hat J}^-
\end{pmatrix}_m\;.
\end{equation}

It is important to remark that the Jones 4-vector $(\bm{\hat J}^+,\bm{\hat J}^-)^T_m$ of Eq.~(\ref{eq:phase-jones}) is evaluated in the $m^\textrm{th}$ medium at the interface of \emph{incidence} onto the $(m+1)^\textrm{th}$ medium (cf.~\cite{He91}, Fig.~4.8), whereas $(\bm{J}^+,\bm{J}^-)^T_m$ is evaluated in the $m^\textrm{th}$ medium at the interface of \emph{emergence} from the $(m-1)^\textrm{th}$ medium. Hence the phase delay between the two definitions corresponds to the thickness of the $m^\textrm{th}$ medium. This is illustrated in more detail in App.~\ref{sec:appA}.

\subsection{Anisotropic and rotated optics} \label{sec:rotations}
In our approach, we want to use the relation (\ref{eq:jonestran4.alt}) to approximately model a train of generally anisotropic optical elements. The main approximation we must impose in order to do so is to assume that the birefringence of any anisotropic material is small enough that the o and e rays, into which the beam entering that material splits, are practically propagating along a common path and with polarizations that remain orthogonal \cite{Ye82}, so 
\emph{the beam can be described by a Jones vector at all time}. 
Then, Eq.~(\ref{eq:jonestran4.alt}) becomes readily applicable, at least under the assumption that the
principal axes of all the optics in the stack are aligned with respect to each other, as well as to the PoI frame in which the $p$ and $s$ components of the field are defined. 

In general, we want to be able to treat a stack of birefringent materials that
are oriented at arbitrary angles $\alpha_m$ from 
the PoI (see Fig.~\ref{fig:perp_geom}),
which requires the
definition of the proper rotation operators necessary to model the transfer across interfaces.
In our treatment, we will always assume that a birefringent optic is a perfeclty 
plane-parallel (i.e., without wedge) uniaxial crystal, which has been cut such that the 
interface between the material and the adjacent medium is either parallel to the optic axis (called ``A-cut'' crystals; see description 
of Fig.~\ref{fig:perp_geom}) or perpendicular to it (called ``C-cut'' crystals).

First of all, let us consider two Jones vectors $\bm{J}$ and $\bm{J'}$ 
connected via a $2{\times}2$ Jones matrix $\mathbf{M}$ 
describing a birefringent optic, $\bm{J'}=\mathbf{M}\bm{J}$. 
Let us further assume for the moment the case of normal incidence. If the optic is rotated---based on the orientation of its principal
axes---by an angle $\alpha$ with respect to the orthogonal axes of 
reference for the two components of the Jones vectors (see Fig.~\ref{fig:perp_geom}), then
\begin{equation} \label{eq:rotation}
\bm{J'}=\mathbf{R}(-\alpha)\mathbf{M}\mathbf{R}(\alpha)\,\bm{J}\;,
\end{equation}
where we introduced the $2{\times}2$ rotation matrix
\begin{equation} \label{eq:defR}
\mathbf{R}(\alpha)=
\begin{pmatrix}
\cos\alpha &\sin\alpha \\
-\sin\alpha &\cos\alpha
\end{pmatrix}\;.
\end{equation}
Equation~(\ref{eq:rotation}) trivially expresses the fact that the mapping by the rotated optic between two Jones vectors must be the same as the mapping by the non-rotated optic between the two counter-rotated Jones vectors, $\bm{{J}'}_\alpha$ and $\bm{J}_\alpha$, i.e.,
\begin{equation} \label{eq:rotated-jones}
\bm{{J}'}_\alpha\equiv\mathbf{R}(\alpha)\bm{{J}'}
=\mathbf{M}\,
\mathbf{R}(\alpha)\bm{J}\equiv\mathbf{M}\bm{J}_\alpha\;.
\end{equation}

If now we consider the problem of Sect.~\ref{sec:formalism}  with the $m$ medium rotated by $\alpha_m$ with respect to the PoI
system of reference,
we can generalize Eq.~(\ref{eq:jonestran4.alt}) to a relation where 
all Jones vectors are expressed in the reference frame of the principal axes of the corresponding medium. 
It is convenient to work with the phase-modified Jones
vectors (\ref{eq:phase-jones}), so we can use
Eq.~(\ref{eq:jonestran4.alt}). As such a relation applies in the
reference frame of the principal axes of the $m^\textrm{th}$ optic rotated to the
angle $\alpha_m$ (see Fig.~\ref{fig:perp_geom}), in analogy with the
Mueller transformation (\ref{eq:rotated-jones}), we immediately arrive at the following \emph{transfer law} between Jones 4-vectors,
\begin{eqnarray} \label{eq:Ye82}
&&
\mathbf{Q}(\alpha_m)
\begin{pmatrix}
\bm{\hat J}^+ \\
\bm{\hat J}^-
\end{pmatrix}_{m-1}
=
\left(
\mathbf{O}_{m-1,m}\,
\mathbf{\Lambda}_m^\ast
\right)
\mathbf{Q}(\alpha_m)
\begin{pmatrix}
\bm{\hat J}^+ \\
\bm{\hat J}^-
\end{pmatrix}_m \nonumber \\
\Rightarrow 
&&
\begin{pmatrix}
\bm{\hat J}^+ \\
\bm{\hat J}^-
\end{pmatrix}_{m-1}
=
\mathbf{Q}(-\alpha_m)\,
\mathbf{O}_{m-1,m}\,
\mathbf{\Lambda}_m^\ast\,
\mathbf{Q}(\alpha_m)
\begin{pmatrix}
\bm{\hat J}^+ \\
\bm{\hat J}^-
\end{pmatrix}_m\;,
\end{eqnarray}
where we defined the $4{\times}4$ rotation matrix
\begin{equation} \label{eq:defQ}
\mathbf{Q}(\alpha)\equiv
\begin{pmatrix}
\mathbf{R}(\alpha) &\mathbf{0} \\
\mathbf{0} &\mathbf{R}(\alpha)
\end{pmatrix}\;,
\end{equation}
and the matrices $\mathbf{\Lambda}_m$ and $\mathbf{O}_{m-1,m}$ 
have the same structures as in Eq.~(\ref{eq:Lmat-Omat}), 
with all quantities referenced to the principal-axes system of the $m^\textrm{th}$ optic.
Since the Fresnel coefficients in the $\mathbf{O}_{m-1,m}$ matrix contain the indexes of refraction of \emph{both} adjacent materials, the indexes of refraction of the preceding $(m-1)^\textrm{th}$ optic must be ``referenced'' (or ``resolved'', using the terminology of \cite{Cl04c}) to the principal-axes frame of the $m^\textrm{th}$ optic, when computing $\mathbf{O}_{m-1,m}$. Such a resolution of the refractive indexes of the $(m-1)^\textrm{th}$ material to the principal axes of the $m^\textrm{th}$ material is an approximate operation of our formalism, which is responsible for the most significant deviations from an exact treatment of the problem \cite{Ha23,Be72}.
The downside is that particular care must be taken in the numerical implementation of the formalism, as described in App.~\ref{app:numerical}; the upside is a demonstrated higher computational efficiency of the approximate approach, when compared with some numerical implementations of the exact theory such as Berreman calculus, while it still provides sufficiently accurate results for quantitative fringe estimation in polarimetric instruments.

\begin{figure}[t!]
\centering
\includegraphics[width=.495\textwidth]{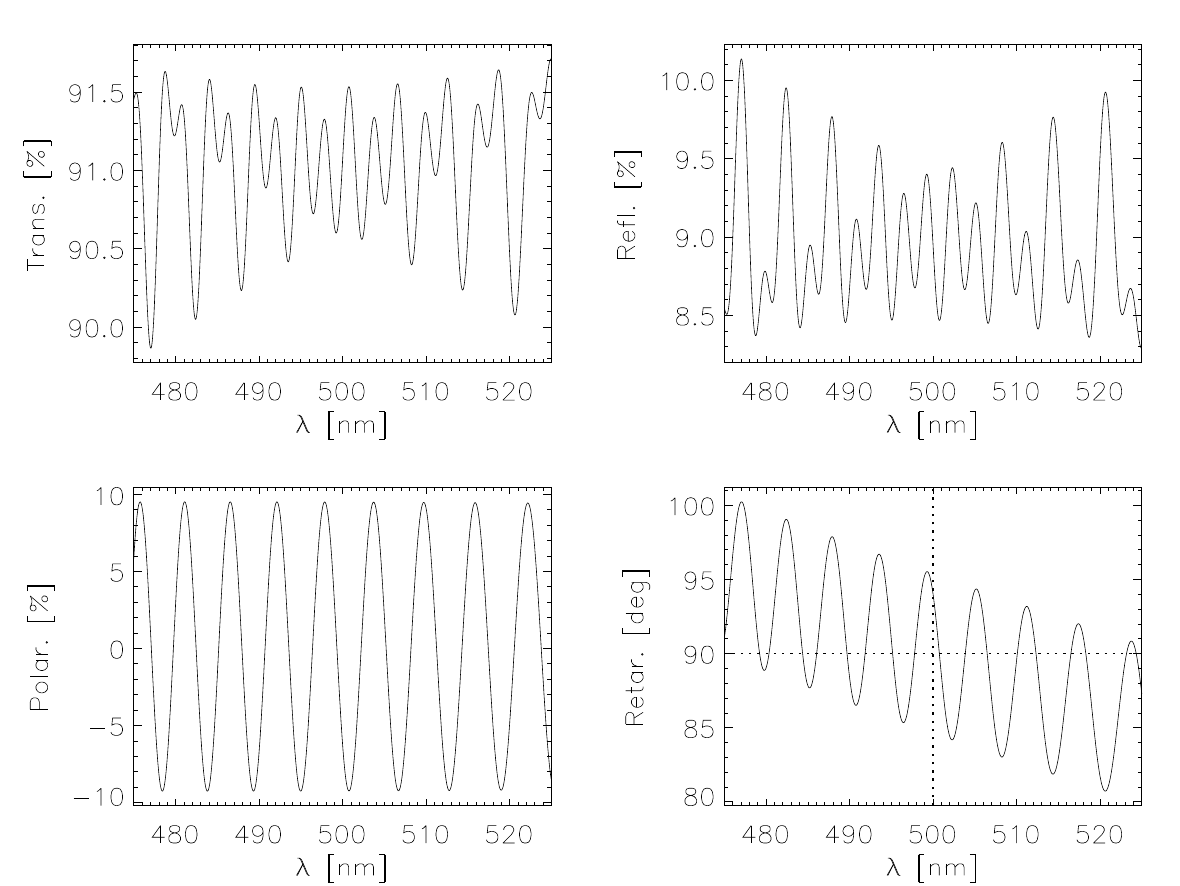}
\includegraphics[width=.495\textwidth]{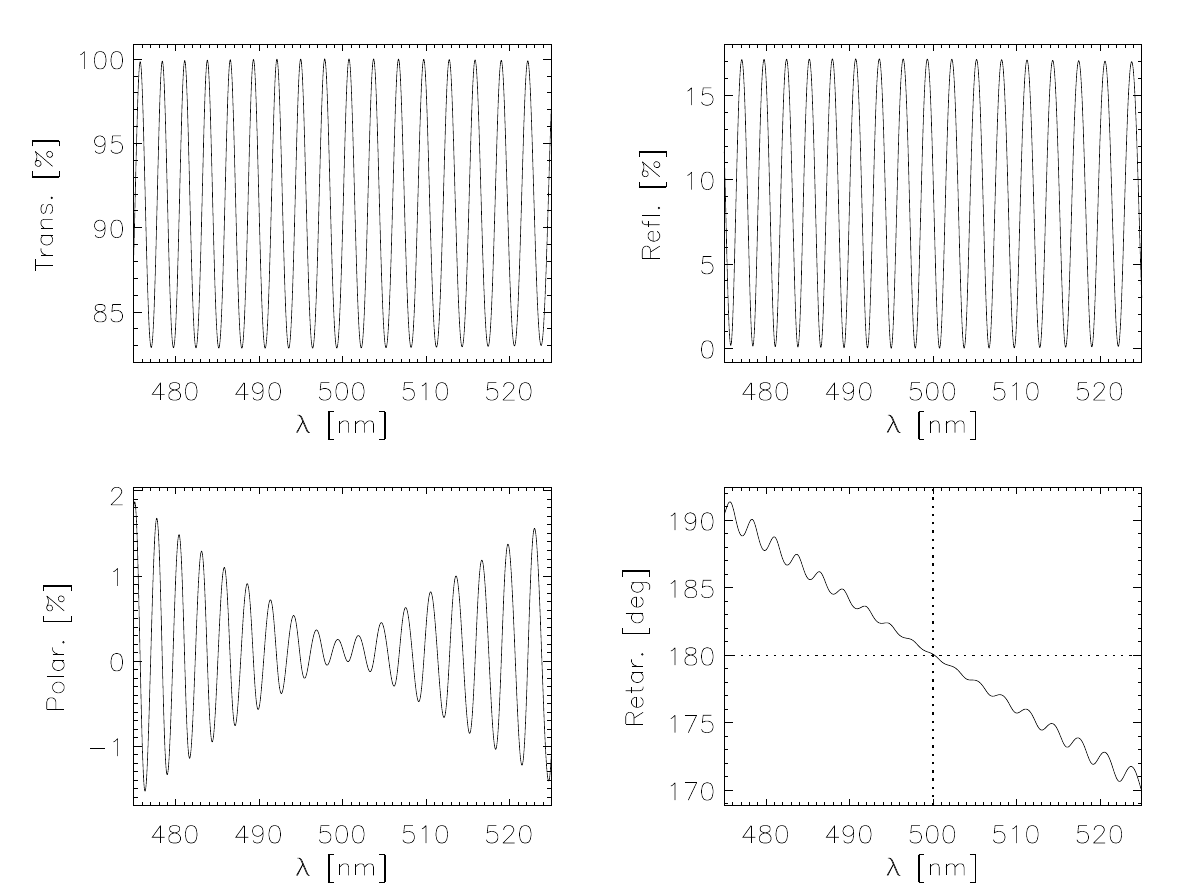}\vspace{5pt}
\includegraphics[width=.495\textwidth]{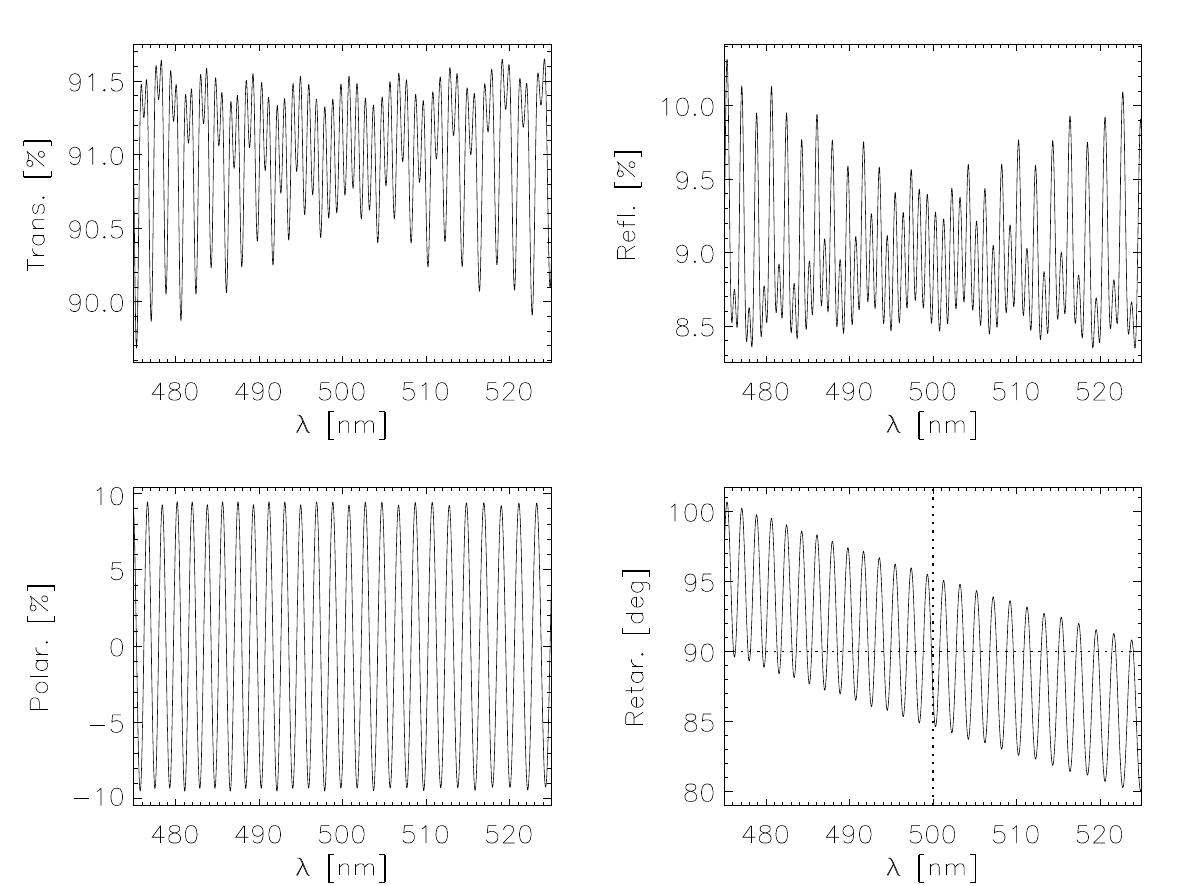}
\includegraphics[width=.495\textwidth]{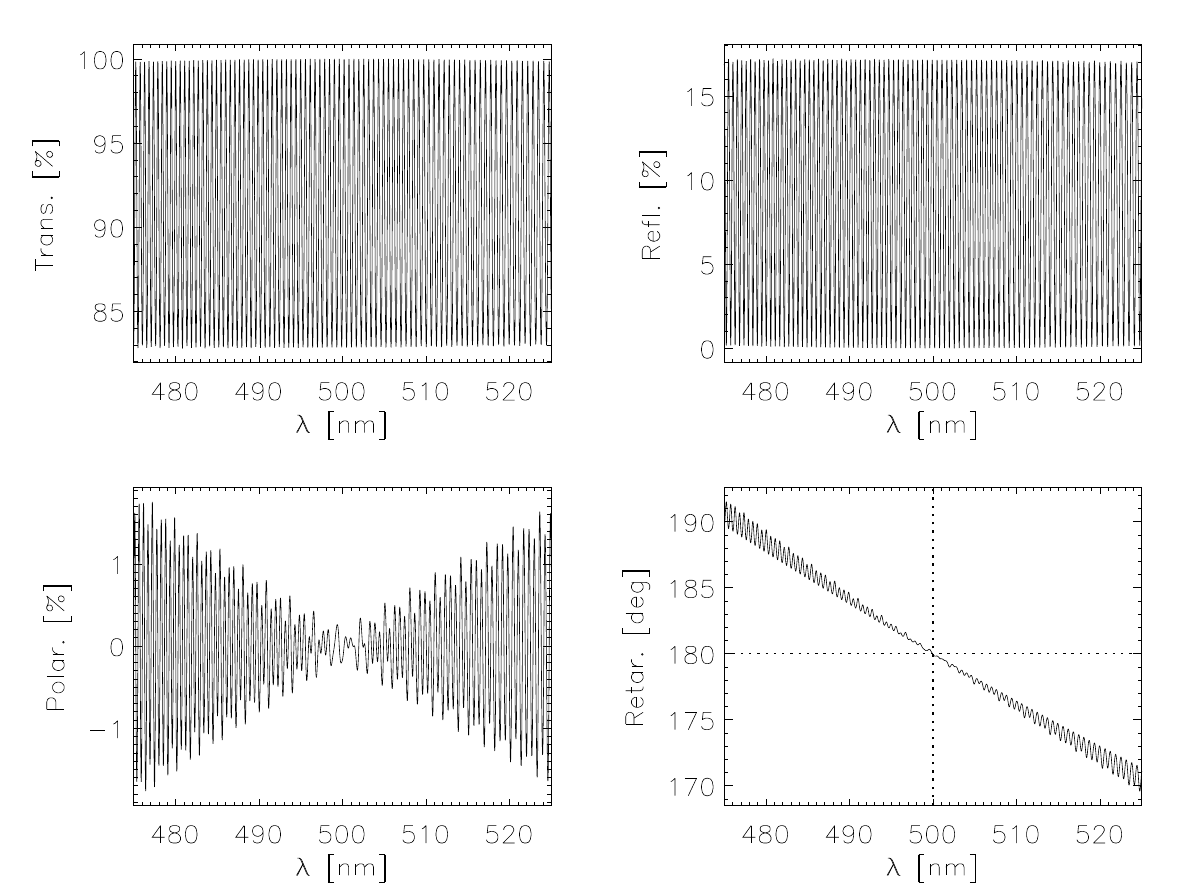}
\caption{Four different examples of birefringent stacks, illustrating their optical and polarization properties, calculated using the formalism presented in this paper. 
The same models were also considered by \cite{Cl04a,Cl04b,Cl04c}. For each example, we show four panels for the
transmittance, reflectance, polarizance, and retardance of the corresponding optical
system, as indicated by the labels. \emph{Top left:} a $\lambda/4$ waveplate at 500\,nm, consisting of
13.5\,$\mu$m of SiO$_2$. \emph{Top right:} a $\lambda/2$ waveplate at the
same wavelength, consisting of 27\,$\mu$m of SiO$_2$. \emph{Bottom
left:} a compound $\lambda/4$ waveplate at the same wavelength, produced 
by combining the two previous waveplates with crossed principal axes.
\emph{Bottom right:} a compound $\lambda/2$ waveplate consisting
of a stack of three $\lambda/2$ waveplates as in the top-right panel, 
followed by a stack of two identical $\lambda/2$ waveplates at
90$^\circ$ from the first stack. In all cases, we assumed a spectral sampling resolution of 50000 and normal incidence.
\label{fig:examples}}
\end{figure}

\begin{figure}[t!]
\centering
\includegraphics[width=.9\textwidth]{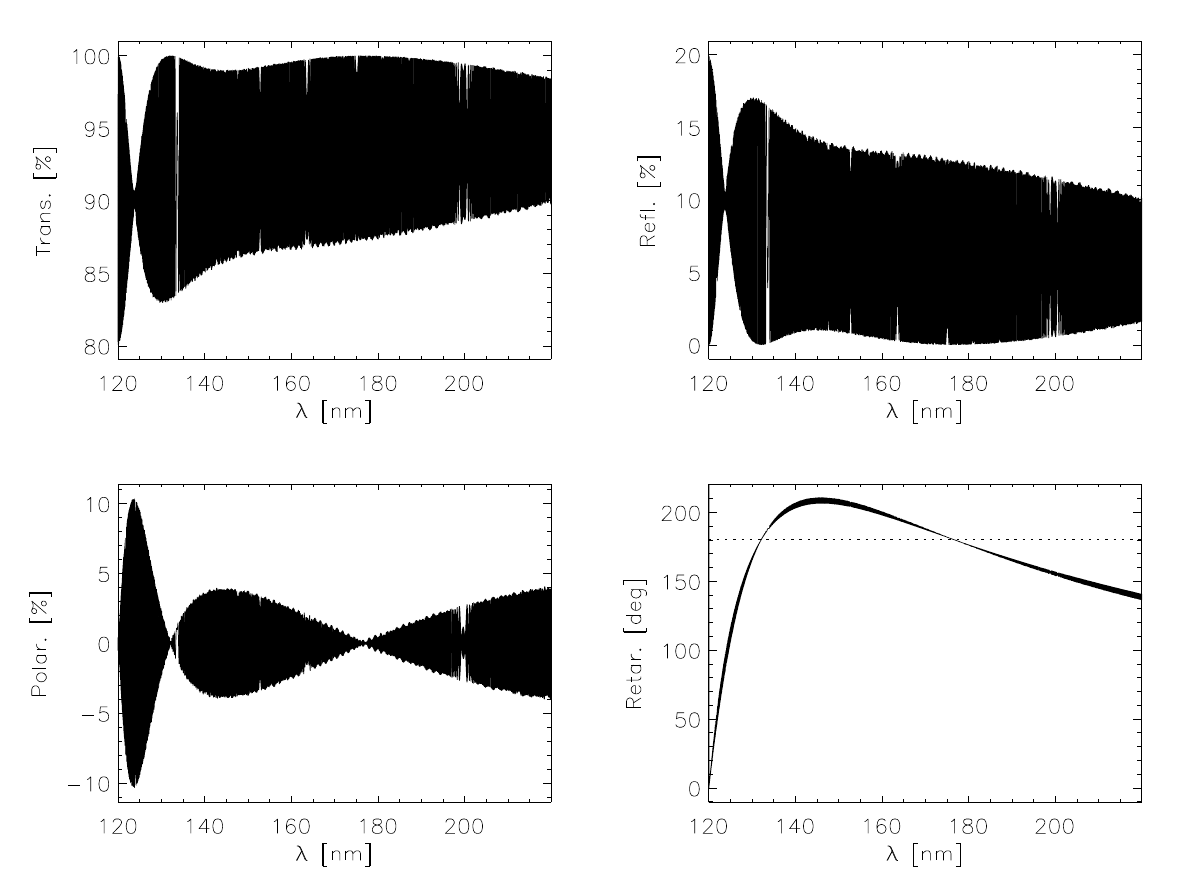}
\caption{Same plotted quantities and experimental conditions as in Fig.~\ref{fig:examples}, but for a MgF$_2$
two-plate compound retarder, optimized to approximately behave as a $\lambda/2$ retarder between 130 and 210\,nm. The two plates have a bias thickness of 800\,$\mu$m.\label{fig:Polstar}}
\end{figure}

From here on, the formalism to model a stack of $N$ birefringent optics
comprised between the $0^\mathrm{th}$ and $(N+1)^\mathrm{th}$ 
materials follows analogously to the recipe 
given by \cite{He91} (cf.~Eq.~4(101)), implying the recursion of
Eq.~(\ref{eq:Ye82}) through the stack of materials, i.e.,
\begin{equation} \label{eq:stack}
\begin{pmatrix}
\bm{\hat J}^+ \\
\bm{\hat J}^-
\end{pmatrix}_{0}
=
\prod_{m=1}^{N+1}\mathscr{M}_m
\begin{pmatrix}
\bm{\hat J}^+ \\
\bm{\hat J}^-
\end{pmatrix}_{N+1}\;,
\quad\mathrm{with}\quad
\mathscr{M}_m\equiv
\mathbf{Q}(-\alpha_m)\,
\mathbf{O}_{m-1,m}\,
\mathbf{\Lambda}_m^\ast\,
\mathbf{Q}(\alpha_m)\;.
\end{equation}

In the following, we explicitly assume that the 
material for both the $0^\textrm{th}$ and $(N+1)^\textrm{th}$ media 
is always the vacuum, so the phase factors associated with those two layers are isotropic, and therefore 
independent of the orientation specified for those media 
(cf.~Eq.~(\ref{eq:phase-jones})).
%
%
%
In addition, because the incident and reflected Jones vectors in the $0^\textrm{th}$ medium are specified at the interface with the $1^\textrm{st}$ medium, there is no phase factor associated with its thickness, so the corresponding phase matrix $\mathbf{\Lambda}_0$ is simply the identity (cf.~Eq.~(\ref{eq:phase-jones})). 
Thus, we can make the following substitutions in Eq.~(\ref{eq:stack})
\begin{displaymath}
\begin{pmatrix}
\bm{\hat J}^+ \\
\bm{\hat J}^-
\end{pmatrix}_{0}
\equiv
\begin{pmatrix}
\bm{J}^+ \\
\bm{J}^-
\end{pmatrix}_{0}\;,
\qquad
\begin{pmatrix}
\bm{\hat J}^+ \\
\bm{\hat J}^-
\end{pmatrix}_{N+1}
\equiv 
\mathbf{\Lambda}_{N+1}
\begin{pmatrix}
\bm{J}^+ \\
\bm{J}^-
\end{pmatrix}_{N+1}\;.
\end{displaymath}
Finally, we impose that no backward propagating beam 
is present in the $(N+1)^\textrm{th}$ material, so 
$\bm{J}^-_{N+1}=0$.
We thus arrive at the final expressions that we must solve in 
order to determine the transmission and polarization properties of the stack,
\begin{equation}
\label{eq:stacksolve.1}
\begin{pmatrix}
\bm{J}^+ \\
\bm{J}^-
\end{pmatrix}_{0}
=
\prod_{m=1}^{N+1}\mathscr{M}_m\,
\mathbf{\Lambda}_{N+1}
\begin{pmatrix}
\bm{J}^+ \\
\bm{0}
\end{pmatrix}_{N+1}\;,
\end{equation}
with $\mathscr{M}_m$ given by Eq.~(\ref{eq:stack}).

As we remarked at the beginning of this section, the former equations have been derived for the case of normal incidence on the stack of optics. When the incoming beam has an incidence
angle $\phi_m>0$ to the surface normal, the rotation angle $\alpha_m$ in the equations above must be replaced by the ``projected'' rotation angle
$\psi_m$ seen by the ray, which is given by
(cf.~\cite{Ye82}, Eqs.~(69))
\begin{equation} 
\label{eq:projected}
\sin\psi_m
=\frac{\sin\alpha_m}{\sqrt{1-\sin^2\phi_m\cos^2\alpha_m}}\;, \quad
\cos\psi_m
=\frac{\cos\phi_m\cos\alpha_m}{\sqrt{1-\sin^2\phi_m\cos^2\alpha_m}}\;,
\end{equation}
or
\begin{equation} \label{eq:psi}
\tan\psi_m=\tan\alpha_m/\cos\phi_m\;,
\end{equation}
where $\alpha_m$ still represents the clocking of the $\bm{o}$-axis of the material 
with respect to the $x$-axis, and $\phi_m$ is the angle of propagation of the beam through 
the material.\footnote{Equations~(\ref{eq:projected}) are easily derived by noting that $\cos\psi_m=\bm{n}\cdot\bm{n}'$, where $\bm{n}$ is the normal to the PoI and $\bm{n}'$ is the normal to the $\langle \bm{o},\bm{k}\rangle$ plane; see Fig.~\ref{fig:perp_geom}.}

We note that $\mathbf{Q}(\psi_m)$ still represents a proper rotation under the small-birefringence approximation \cite{Ye82}, as the polarizations of the o and e rays can be considered to remain orthogonal \emph{regardless of the propagation direction}, and thus they can still be represented as the components of a Jones vector propagating in a specific direction. A more in depth discussion is given in App.~\ref{app:numerical}, where we also present a first-order relaxation of this approximation that enables an improved quantitative description of the effects of birefringence in optical systems in finite f/\# beams. An immediate consequence of this is that the propagation angle $\phi_m$, and therefore the projected rotation angle $\psi_m$ as well, can be different for the o and e rays, in which case $\mathbf{Q}(\psi_m)$ is no longer strictly orthogonal.

\begin{figure}[t!]
    \centering
\includegraphics[width=\linewidth]
    {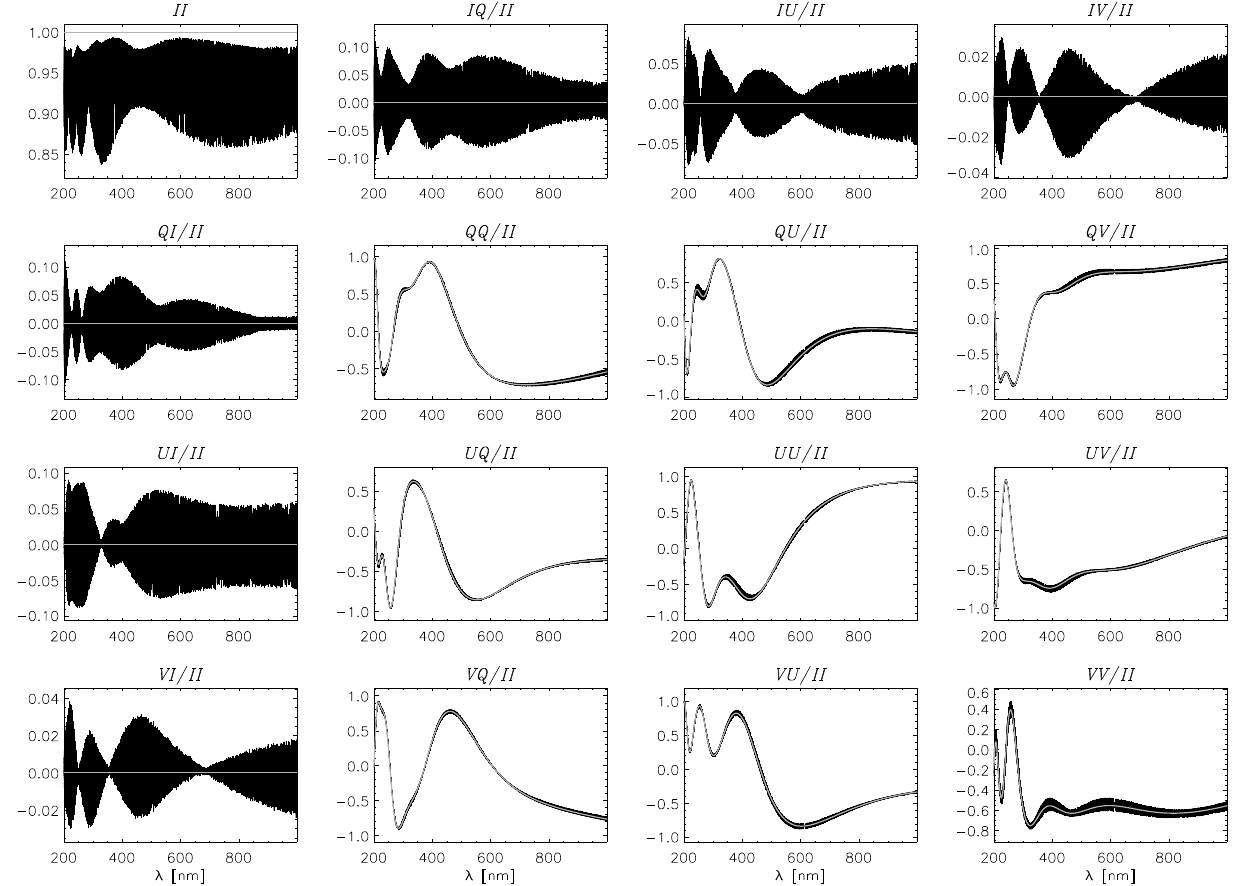}
    \caption{Wavelength dependence between 200 and 1000\,nm of the (intensity normalized) Mueller matrix of a polychromatic modulator (PCM) optimized for full-Stokes polarimetry between 400 and 1000\,nm. The polarization fringes are calculated  with a spectral resolution of 20000. The PCM design uses three compound retarders in the configuration MgF$_2$-SiO$_2$-MgF$_2$, where the two MgF$_2$ elements are identical. The gray curves plotted over the fringes represent the ideal Mueller matrix from the PCM design.} 
    \label{fig:check}
\end{figure}

In the case of a conducting medium with complex index of refraction $\nu$, the quantities necessary to evaluate the proper Fresnel coefficients (see App.~\ref{app:numerical}, Eqs.~(\ref{eq:Fresnel}--\ref{eq:Fresnel.last})) satisfy the generalized Snell's law (see, e.g., Eq.~4(33) of \cite{He91}; also App.~\ref{app:numerical} in this paper),
\begin{equation} \label{eq:snell}
\nu_m\sin\phi_m=\nu_{m-1}\sin\phi_{m-1}\;,
\end{equation}
where the propagation angles $\phi_{m-1}$ and $\phi_m$ generally become complex quantities.
In such a case, the true (real) propagation angles through the conducting medium, which must be used to evaluate Eqs.~(\ref{eq:projected}), do \emph{not} correspond to the (complex) solution of Eq.~(\ref{eq:snell}), and must instead be determined separately (see App.~\ref{app:numerical}; also App.~XV~A of \cite{Di63}).

Figures \ref{fig:examples} and \ref{fig:Polstar} show  examples of various optical and polarization properties of different stacks of birefringent optics, modeled with the
formalism presented above. In particular, the results of Fig.~\ref{fig:examples} can be directly compared with those presented in \cite{Cl04a,Cl04b}.

A further validation of the formalism is provided by comparing the predicted wavelength-dependent Mueller matrix with the one obtained in the ideal case when wave interference and Fresnel losses are neglected.
Figure~\ref{fig:check} shows the example of a ``polychromatic'' modulator (PCM \cite{To10}) consisting of a MgF$_2$-SiO$_2$-MgF$_2$ 6-plate stack, and optimized for full-Stokes polarimetry between 400 and 1000\,nm. The ideal Mueller matrix of this design is represented by the gray curves.
We note how the ideal solution perfectly overlaps with the fringe patterns calculated with our formalism, with the obvious exception of the transmittance element $II$, which is affected by Fresnel losses that have not been normalized in this plot. 

All these figures demonstrate the close connection among intensity, retardance, and polarizance/diattenuation fringes. This is clearly manifested at specific conditions of constructive and destructive interference of the transmitted and reflected waves, which is attained for $m\pi/2$ values of the retardance, with $m$ an integer. In particular, Fig.~\ref{fig:examples} shows how retardance and polarizance/diattenuation fringes are suppressed around half-wave conditions, while intensity fringes, in both transmission and reflection, tend to be suppressed around quarter-wave conditions. This is also clearly demonstrated by Fig.~\ref{fig:Polstar}, where the occurrence of nulls and maxima of the fringe amplitudes in intensity and polarizance/diattenuation across the spectrum appear to be in ``opposition of phase'', with the two nulls of polarizance fringes specifically occurring at the two wavelengths where the compound retarder is exactly half-wave.

\subsection{The effect of finite f/\#}

\begin{figure}[t!]
    \centering
\includegraphics[width=\linewidth,
    clip]{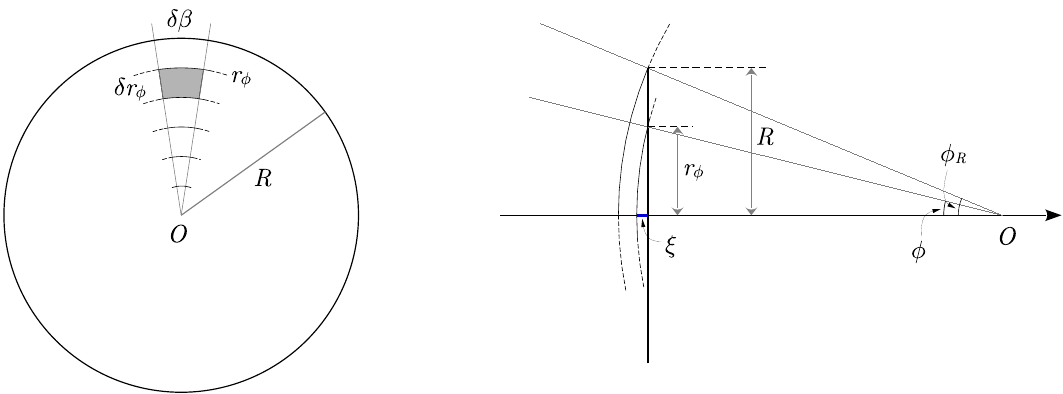}
    \caption{Geometric constructs to determine the elemental surface (left) and phase delay (right) pertaining to a ray incident with an angle $\phi$ on the stack of optics of radius $R$, assuming a spherical wavefront with $\mathrm{f/\#} = (2\tan\phi_R)^{-1}$.}
    \label{fig:shapedbeam}
\end{figure}

As pointed out in the previous section, the formalism we propose can also be applied to the case of a ray pencil with an incidence angle $\phi$ with the surface normal of the stack different from zero. This allows us to extend the application of this formalism to the case of shaped beams, in particular, for a converging or diverging spherical wavefront produced by an ideally stigmatic optical system, in order to model the spatial-dependent fringes (such as Haidinger's fringes; see, e.g., \cite{Di63}, \S5.17) that are produced on an image plane with such beam configurations. Estimation of fringe amplitudes in real astronomical instruments must account for averaging over these spatial fringes \cite{JATIS2, JATIS11}, as empirically verified in \cite{JATIS3}.

Two quantities need to be evaluated in order to do so (see Fig.~\ref{fig:shapedbeam}): 1) the elemental solid angle associated with the ray direction $(\phi,\beta)$ and angular steps $(\delta\phi,\delta\beta)$, where $\phi$ is the angle of incidence and $\beta$ is the azimuth of the PoI around the optical axis of the system (assumed for simplicity to be normal to the surface of the stack), counted from the same zero reference as for the rotation matrices $\mathbf{Q}(\alpha)$ introduced in the previous section; and 2) the phase delay associated with the spherical wavefront crossing the stack of optics at different distances $r_\phi$ from the optical axis. In order to derive our results, we assume that the optic stack is always fully filled by the incoming beam.

\begin{figure}[t!]
    \centering
\includegraphics[width=\linewidth]{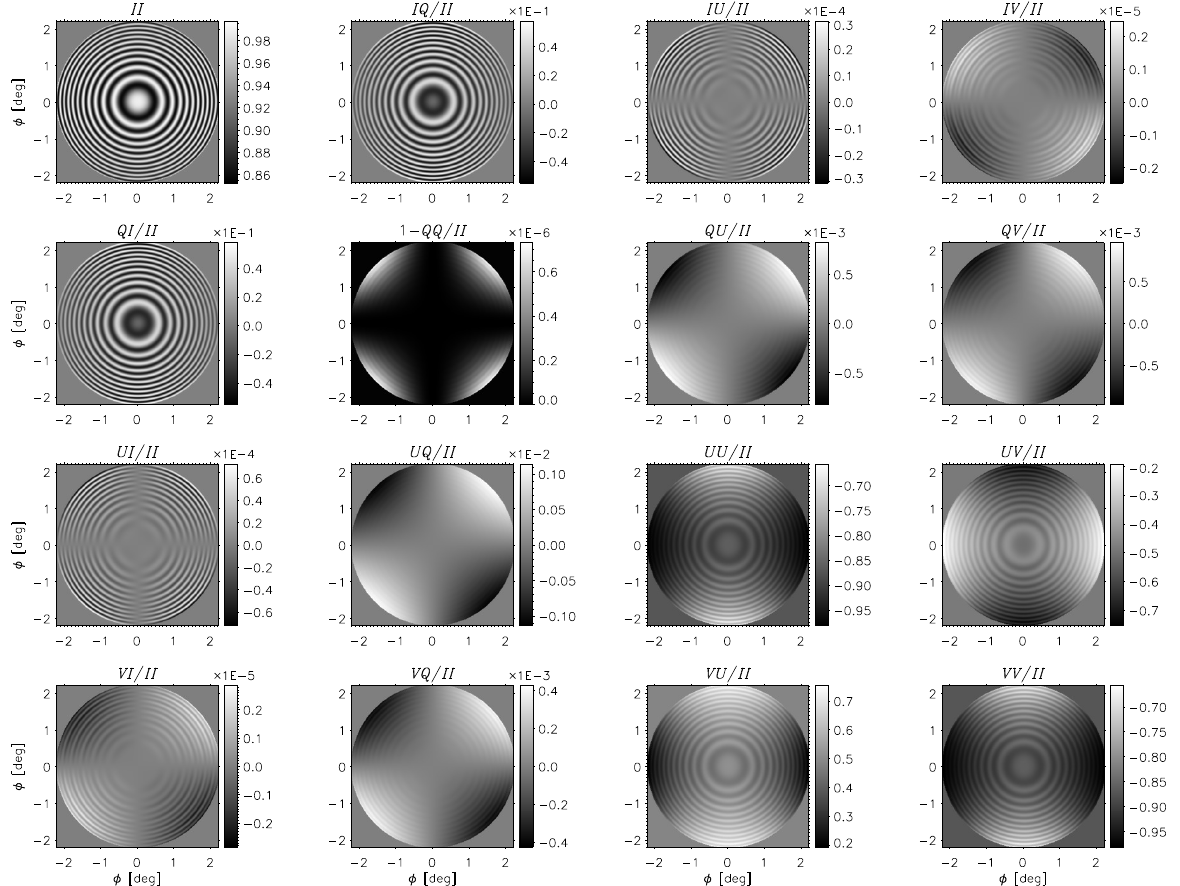}
    \caption{Example of spatial-dependent polarization fringes in monochromatic light, as a function of the incidence angle $\phi$ and azimuth $\beta$, for each of the 16 Mueller matrix elements (intensity normalized) of the near-$\lambda/2$ MgF$_2$ compound retarder of Fig.~\ref{fig:Polstar}. The fringes are evaluated at 146\,nm, around which the waveplate retardance differs the most from the target $\lambda/2$ within the design spectral range, as demonstrated by the range of values taken by the $UU$ and $VV$ elements. The adopted half-cone aperture is $2.2^\circ$, corresponding to a f/13 beam. In these plots and the following, the reference direction for positive $Q$ and for the orientation of the stack ($x$-axis in Fig.~\ref{fig:perp_geom}) is vertical.}
    \label{fig:spatial_fringes}
\end{figure}

\begin{figure}[t!]
    \centering
\includegraphics[width=\linewidth]{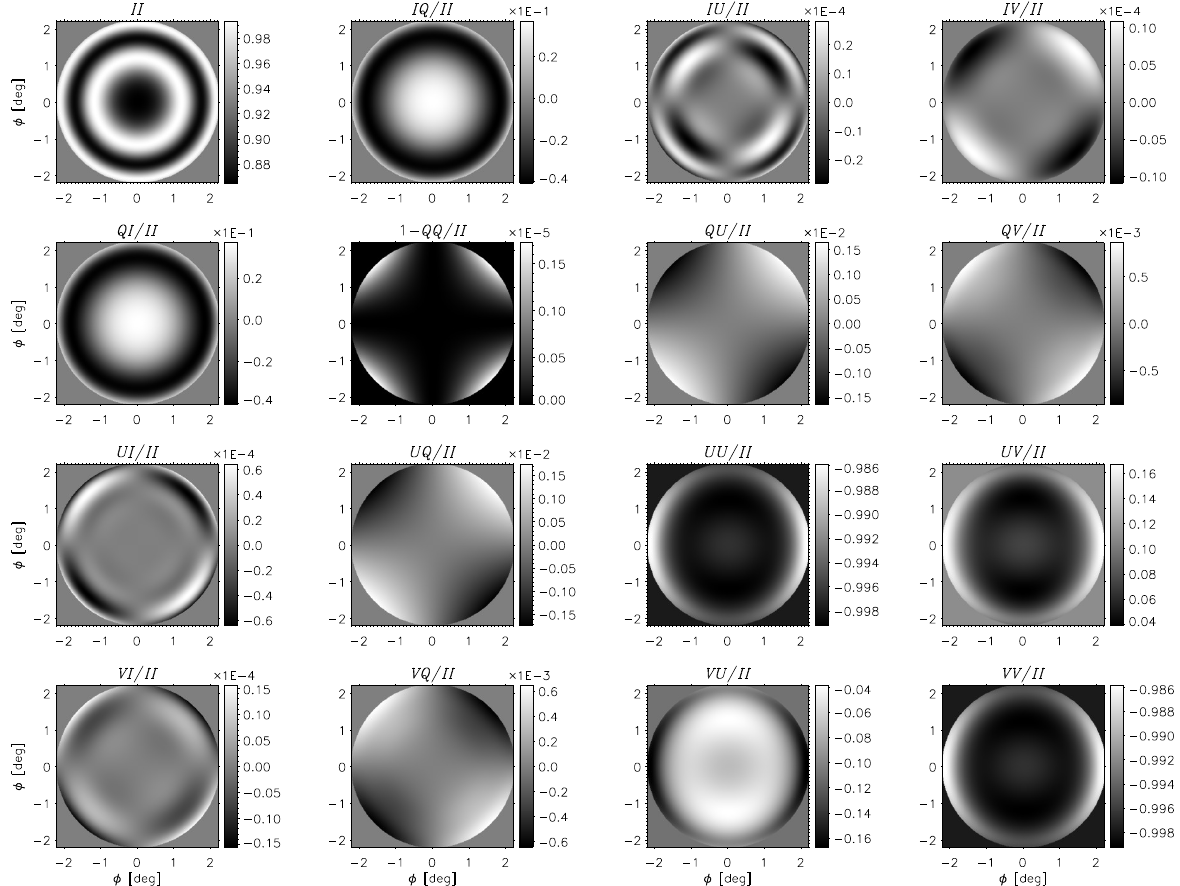}
    \caption{Same as Fig.~\ref{fig:spatial_fringes}, but for a $\lambda/2$ achromat retarder consisting of a  SiO$_2$-MgF$_2$ compound, optimized between 400 and 600\,nm. The fringes are evaluated at 500\,nm, where the achromat has nearly ideal $\lambda/2$ behavior. Like for Fig.~\ref{fig:spatial_fringes}, we assumed a cone aperture corresponding to a f/13 beam.}
    \label{fig:spatial_fringes_achro}
\end{figure}

\begin{figure}[t!]
    \centering
    \includegraphics[width=\linewidth]
    {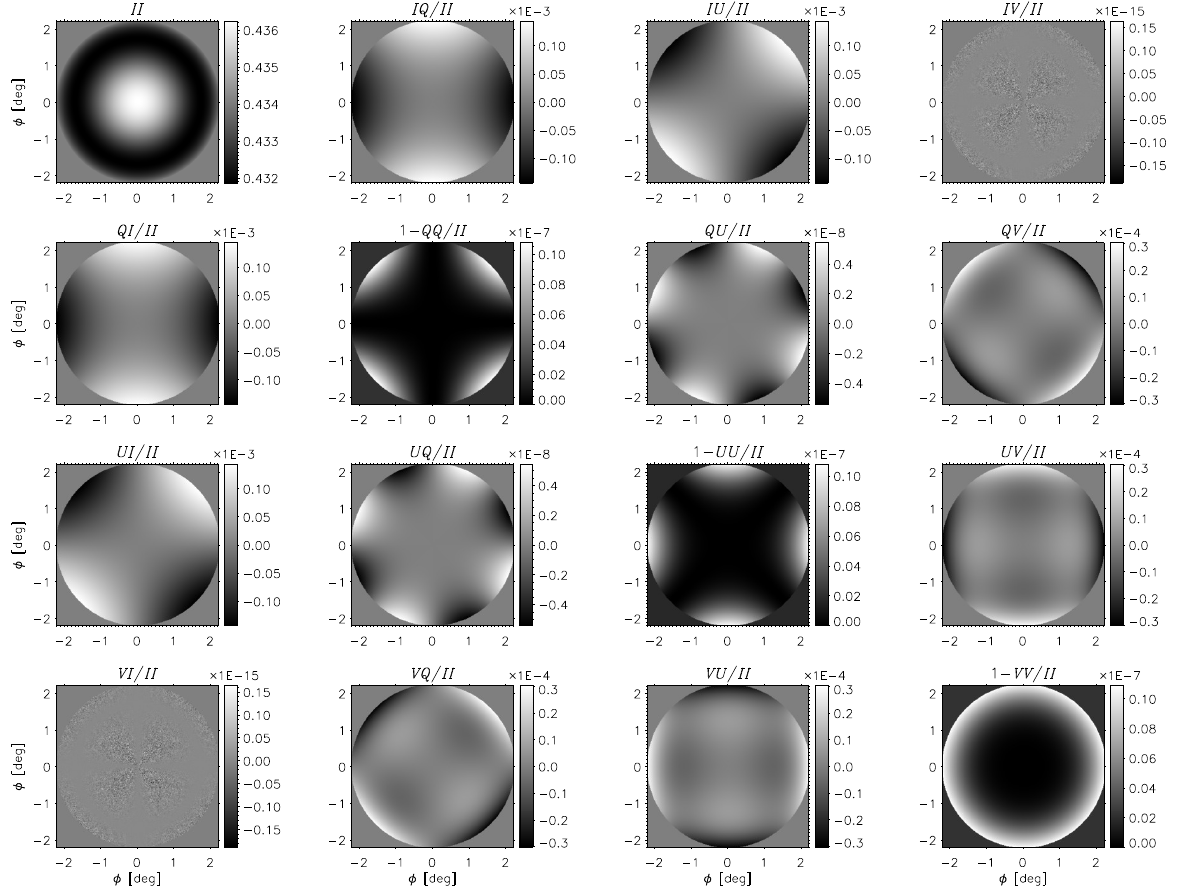}
    \caption{
    Same as Fig.~\ref{fig:spatial_fringes}, but for
    an absorptive plate of fused silica (amorphous SiO$_2$) with a thickness of 1.1\,mm, illuminated by a f/13 beam of monochromatic light at 4.5\,$\mu$m wavelength. For the complex refractive index at this wavelength we used estimations from vendor provided data, with $n=1.3677$ and $k=2.5\times 10^{-4}$.}
    \label{fig:absorb}
\end{figure}

The elemental surface shown in the left panel of Fig.~\ref{fig:shapedbeam} is given by
\begin{eqnarray*}
\delta S(\phi,\beta)
    &=&\pi\Bigl[r_\phi^2-(r_\phi-\delta r_\phi)^2\Bigr] \frac{\delta\beta}{2\pi}
    =\pi R^2\,\frac{r_\phi^2-(r_\phi-\delta r_\phi)^2}{R^2}\,\frac{\delta\beta}{2\pi }\\
    &=&\pi R^2\,\frac{\tan^2\phi-\tan^2(\phi-\delta\phi)}{\tan^2\phi_R}\,\frac{\delta\beta}{2\pi}\;,
\end{eqnarray*}
and thus the corresponding elemental solid angle is
\begin{equation}
\delta\Omega(\phi,\beta)=\frac{\delta S(\phi,\beta)}{\pi R^2}=\frac{\tan^2\phi-\tan^2(\phi-\delta\phi)}{\tan^2\phi_R}\,\frac{\delta\beta}{2\pi}\;.
\end{equation}
%
%
%

The phase delay of the ray incident at $\phi$ with respect to the ray at normal incidence is $\psi_\phi=2\pi\,\xi/\lambda$, where $\lambda$ is the  wavelength, and the segment $\xi$ is highlighted in the right panel of Fig.~\ref{fig:shapedbeam}. From that geometry, we easily derive
$\xi=R\,(\sec\phi-1)/\tan\phi_R$,
%
%
hence
\begin{equation}
\psi_\phi=
2\pi\,\frac{R}{\lambda}\,\frac{\sec\phi-1}{\tan\phi_R}\;.
\end{equation}
Accordingly, the Jones matrices derived from Eq.~(\ref{eq:stacksolve.1}) must be multiplied by the phase factor $\exp(\imag \psi_\phi)$, for each of the $\phi$ angles sampling the ray fan of the incoming beam.

Figures~\ref{fig:spatial_fringes} and \ref{fig:spatial_fringes_achro} show two examples of the angular dependence of polarization fringes for each element of the $4\times4$ Mueller matrix of
a stack of birefringent optics. 
Figure~\ref{fig:spatial_fringes} corresponds to the near-$\lambda/2$ retarder of Fig.~\ref{fig:Polstar}, evaluated at $\lambda=146$\,nm, where the retardance behavior is the farthest from the $\lambda/2$ target over the spectral range of optimization of the compound design. These fringes vary rapidly with wavelength, so their amplitude at the focal plane may be significantly reduced by the smearing produced by the instrument PSF, depending on its spectral resolution. 
Figure~\ref{fig:spatial_fringes_achro} shows the case of an achromatic $\lambda/2$ retarder consisting of 420\,$\mu$m of SiO$_2$ crossed with 348.6\,$\mu$m of MgF$_2$. This design is inspired by the achromat proposed by \cite{Cl04c}, optimized between 400 and 600\,nm. The fringes are evaluated at 500\,nm, where the retardance of the achromat is close to ideal in normal incidence, as demonstrated by the range of values of the $UU$ and $VV$ elements. This is better seen if we normalize the $UU$ and $VV$ elements by the $II$ element, in which case such normalized elements are all very close to $-1$.

It is important to remark that the effect of the phase delay $\psi_\phi$ cancels out when calculating spatially resolved Mueller-matrix maps of the type shown in Figs.~\ref{fig:spatial_fringes} and \ref{fig:spatial_fringes_achro}, and therefore it can simply be neglected for such a purpose.

\subsection{Absorptive materials}


\begin{figure}
\centering
\includegraphics[width=.75\textwidth]{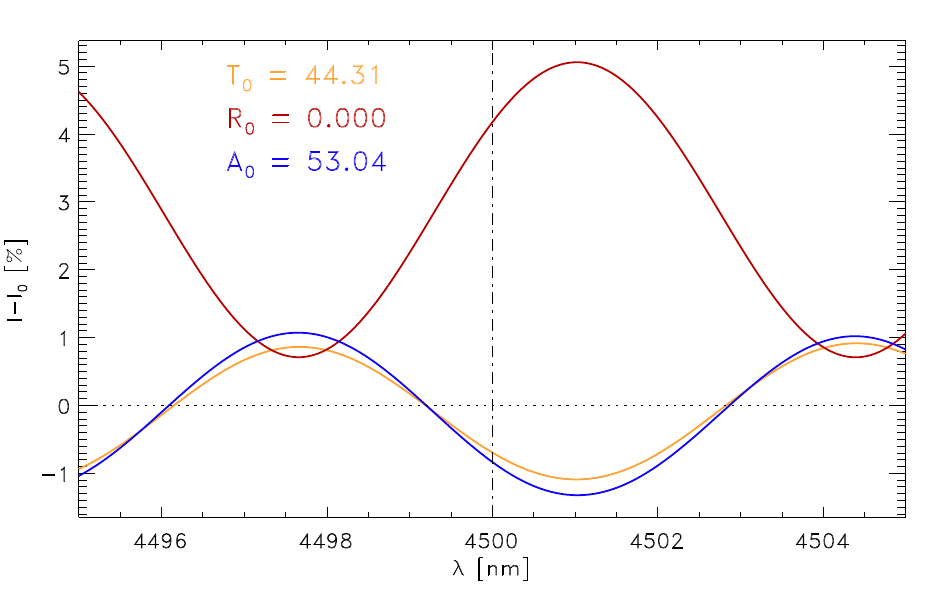}
\caption{\label{fig:fringes_fs_calpol1_4500nm_absorptive}  Calculation of intensity fringes for normal incidence on a 1.1\,mm thick fused-silica window, as employed in DKIST CalPol1 \cite{JATIS8}. Reflectance fringes (red curve) oscillate between 
0.7\% and 5.1\%,
transmittance fringes (orange curve) between 
43.2\% and 45.2\%, and absorbance fringes (blue curve)
between 51.7\% and 54.1\%. The numbers specified inside the plot are the offset values at the marked zero reference, from which the plotted curves are calculated.}
\end{figure}

We conclude the presentation of our formalism by showing how it can account for the behavior of fringe interference in non-transparent media. As mentioned earlier, this implies that the refractive index of the medium is a complex number, but nonetheless the formalism naturally extends to such a case.  Figure~\ref{fig:absorb} shows the spatial fringes for an absorptive fused-silica (FS) window in air, with a thickness of 1.1\,mm, and refractive index $n=1.3677$. The extinction coefficient $k$ was set to 2.5$\times10^{-4}$ to match typical transmittance curves of FS optical glass in the literature. This model corresponds to the DKIST calibration polarizer substrate \cite{JATIS8}, and our formalism estimates an average transmittance of 44.3\% around 4500\,nm, with an average reflectance of 2.9\%. The reflectance fringes oscillate between approximately 0.7\% and 5.1\%, hence never realizing a condition of zero reflectance, which is instead always attained in the case of a non-absorptive plate at the proper wavelengths.

It is instructive to test numerical models of fringes in isotropic absorptive plates against simple first-order approximations that can be derived directly from first principles. For simplicity, we consider the model of Fig.~\ref{fig:absorb} for the case of normal incidence, so the Fresnel coefficients at the interface between two different materials only depend on the (complex) indexes of refraction. 

A field of unit amplitude incident on the plate is split into a reflected and a transmitted beam at the interface, minus any absorptive losses. The minimal model to estimate interference fringes in the reflectance of the plate is to consider the superposition of the first and second reflection.  These have relative field amplitudes $\mathcal{R}_0=r$ and $\mathcal{R}_1=tr't'\exp(-\imag 2\delta)$, respectively, where $r,t$ and $r',t'$ are Fresnel coefficients at the two interfaces, and $\delta$ is given by Eq.~(\ref{eq:delta}) (see Sect.~4.4 of \cite{He91}, in particular Fig.~4.3). From the Fresnel equations (\ref{eq:Fresnel}--\ref{eq:Fresnel.last}), we see that $r'=-r$, while $\mathcal{T}_0=tt'\exp(-\imag\delta)\equiv\nu t^2\exp(-\imag\delta)$ represents the lowest-order approximation for the amplitude of the transmitted field through the plate. 
Therefore, the first-order approximation for the reflected field is
\begin{displaymath}
\mathcal{R}\approx\mathcal{R}_0+\mathcal{R}_1=r\bigl(1+\mathcal{T}_0\,\mathrm{e}^{-\imag\delta}\bigr)
\equiv r\bigl[1+|\mathcal{T}_0|\,\mathrm{e}^{-\gamma}\,\mathrm{e}^{-\imag(\beta+\chi)}\bigr]\;,
\end{displaymath}
where for the last equivalence we set $\delta=\beta-\imag\gamma$ to account for the complex nature of the refractive index $\nu$, and we also adopted the polar form of the complex amplitude $\mathcal{T}_0=|\mathcal{T}_0|\,\mathrm{e}^{-\imag\chi}$, with $\chi$ a real number. 
When $k\ll n$, as in the model of Fig.~\ref{fig:absorb}, it can be shown that $|\mathcal{T}_0|^2\approx|\mathcal{T}_0|\,\mathrm{e}^{-\gamma}$, so the previous approximation for the reflected field becomes
\begin{equation}
\mathcal{R}\approx r\bigl[1+|\mathcal{T}_0|^2\,\mathrm{e}^{-\imag(\beta+\chi)}\bigr]\;.
\end{equation}
Accordingly, the first-order approximation of the reflectance is
\begin{equation} \label{eq:approxfull}
|\mathcal{R}|^2\approx
|r|^2
\bigl[1+|\mathcal{T}_0|^4
+2\,|\mathcal{T}_0|^2\cos(\beta+\chi)\bigr]\;,
\end{equation}
leading to the expressions $|r|^2
(1\pm|\mathcal{T}_0|^2)^2$ for the maxima and minima of the reflectance, with
a peak-to-valley (PV) fringe amplitude
\begin{equation} \label{eq:approx}
I_{\rm PV}\approx 
4\,|r|^2 |\mathcal{T}_0|^2\;.
\end{equation}

Figure~\ref{fig:fringes_fs_calpol1_4500nm_absorptive} plots the intensity fringes computed with our formalism for a beam in normal incidence on the absorptive FS plate of Fig.~\ref{fig:absorb}. For that model we have $|r|^2\approx 2.4\%$ and $|\mathcal{T}_0|^2\approx 44.3\%$, thus $I_{\rm PV}\approx 4.25\%$, which closely matches the PV amplitude of the reflectance curve in Fig.~\ref{fig:fringes_fs_calpol1_4500nm_absorptive}. We remark that the approximate maxima and minima of the reflectance fringes given above yield $5.0\%$ and $0.7\%$, which also match quite closely the values in Fig.~\ref{fig:fringes_fs_calpol1_4500nm_absorptive}. In particular, Eq.~(\ref{eq:approxfull}) demonstrates why the minimum reflectance at the interface with an absorptive material never reaches zero, unlike the case of a transparent medium ($|\mathcal{T}_0|^2\approx 1$).

The approximation (\ref{eq:approx}) can be generalized to the case of a uniform plate between semi-infinite materials of different refractive indexes $\nu_1$ and $\nu_2$ (e.g., an AR coating of index $\nu$ deposited on top of a substrate). Its derivation follows the same steps as in the previous example (see Sect.~4.4 of \cite{He91}), where now $\mathcal{R}_0=r_1$ and $\mathcal{R}_1=t_1 r_2 t_1'\exp(-\imag 2\delta)$. Using similar algebraic manipulations as in the former example leads to the expressions $(|r_1|\pm|r_2||\mathcal{T}_1|^2)^2$ for the maxima and minima of the reflectance fringes, and therefore
\begin{equation}
I_{\rm PV}\approx 4\,|r_1||r_2|\,|\mathcal{T}_1|^2\;.
\end{equation}
Here $|\mathcal{T}_1|^2$ is the transmittance of the plate \emph{as if it were fully immersed in the medium of index $\nu_1$}, whereas $r_1$ and $r_2$ are the reflectivity amplitudes at the two interfaces of the true system.

To conclude this section, we want to comment on the sensitivity of fringes to the extinction coefficient $k$ of absorptive materials. The values of $k$ range by orders of magnitude across wavelength, they typically are affected by larger experimental errors than in the case of the real index $n$, especially for moderately absorptive materials, and they also can vary quite widely between different material grades. At the end of Sect.~\ref{sec:rotations} we commented on the interdependence of intensity and polarization fringes, and how this manifests particularly around specific values of the retardance. Thus, it can be expected that both intensity and polarization fringes will generally be impacted by the absorption characteristics of a material, and how uncertainties in the value of $k$ may impact the modeling accuracy of interference fringes is an important question to address. Our finding is that, by measuring optical transmission with the typical accuracy of commonly available tools, as in the case of the DKIST polarizer-window example of Fig.~\ref{fig:fringes_fs_calpol1_4500nm_absorptive}, we are able to constrain the actual $k$ value precisely enough for fringe modeling purposes. 
In summary, when absorption values are large enough to impact fringe characteristics, the corresponding $k$ values are sufficiently well constrained by standard optical transmission measurements
for accurately modeling both intensity and polarization fringes.  
On the other hand, for cases with very small and often highly uncertain $k$ values, the impact of absorptivity on fringes is normally negligible.



\section{The impact of polarization fringes on optical-design trades}

Using the modeling framework derived in the previous section, we now want to illustrate how the ability to model polarization fringes in layered materials can inform the design choices for polarimetric instrumentation. In particular, we want to show explicitly how the choices of material thicknesses (including those of any non birefringent media in the stack) and beam shapes can be used to mitigate the impact of fringes on polarimetric accuracy. In order to do so, we analyze the behavior of PCMs optimized to work between some prescribed spectral ranges with a required spectral resolution.

One commonly encountered design choice is that of placing polarization modulators in collimated light in order to minimize the field-of-view (FOV) dependence of the birefringence of the optics. On the other hand, the manifestation of polarization fringes is amplified in such a configuration, and the negative impact of strong polarization artifacts in science data is typically a much worse condition to face in practice, as the possibility to remove such artifacts through post-processing of the data is often rather limited (see, e.g., \cite{CL19} and references therein). For this reason, a trade study is always recommended to identify the slowest light beam through a birefringent optic compatible with the design, for which the amplitude of polarization fringes is suppressed below a given target.

Another design choice that is commonly impacted by the manifestation of polarization fringes is whether to adopt air- or oil-gapped optical assemblies, which offer lower fabrication costs and risks, or to follow the more expensive option of molecular adhesion (aka, optical contacting) at one or more of the materials' interfaces, in order to minimize the number and amplitude of the Fresnel reflections that interfere with the transmitted wavefront.

\begin{figure}[t!]
    \centering
\includegraphics[width=\linewidth]{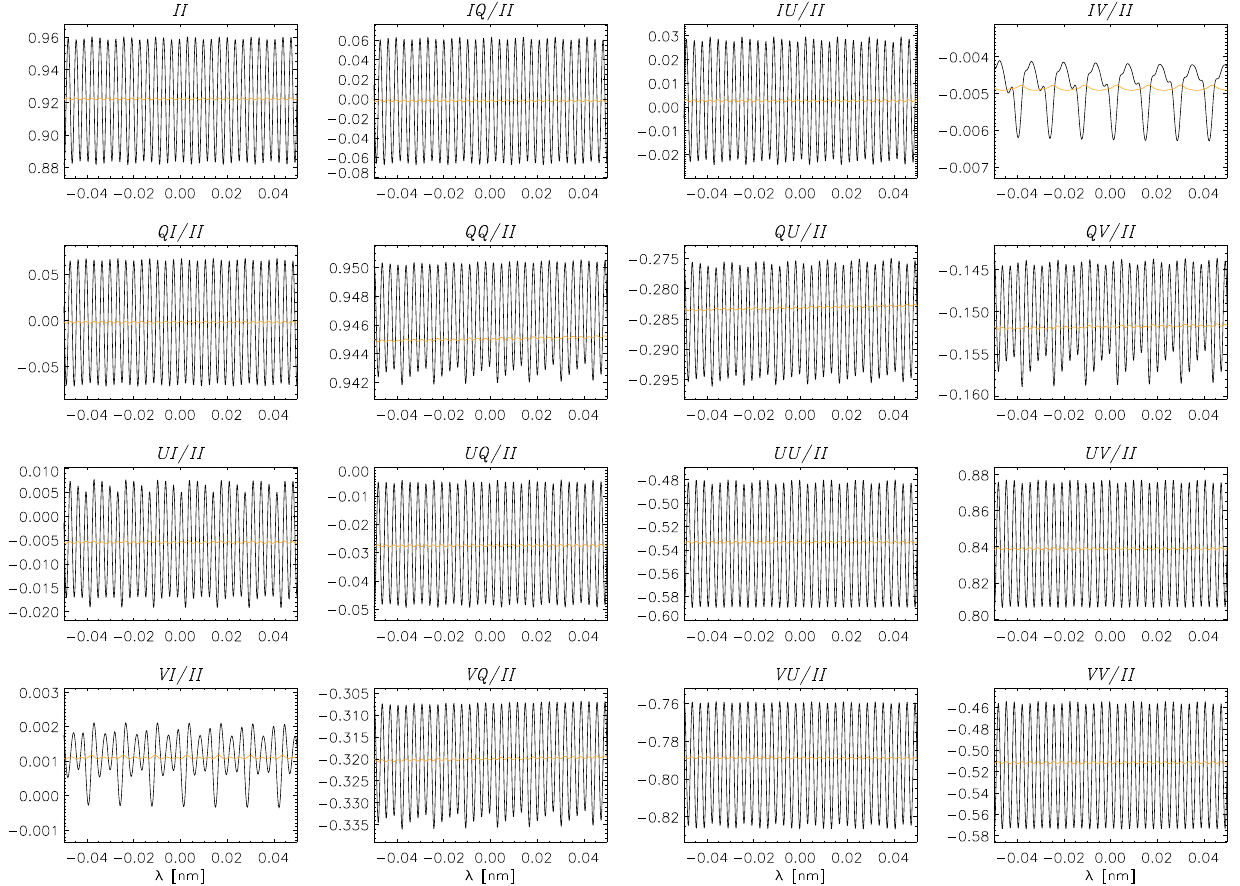}
    \caption{Polarization fringes in the Mueller matrix (intensity normalized) of an optically contacted MgF$_2$ PCM, optimized between 125 and 285\,nm, according to the recipe in Table~\ref{tab:PCM}. The adopted spectral resolution is such that the fringes are fully resolved. A narrow spectral region about 0.1\,nm wide around 144\,nm is shown here, spanning several fringe periods. The black curves show the case of the modulator illuminated by a collimated beam in normal incidence, while the orange curves show the same modulator illuminated by a f/13 beam.}
    \label{fig:Polstar PCM}
\end{figure}

\begin{figure}[t!]
    \centering
\includegraphics[width=\linewidth]{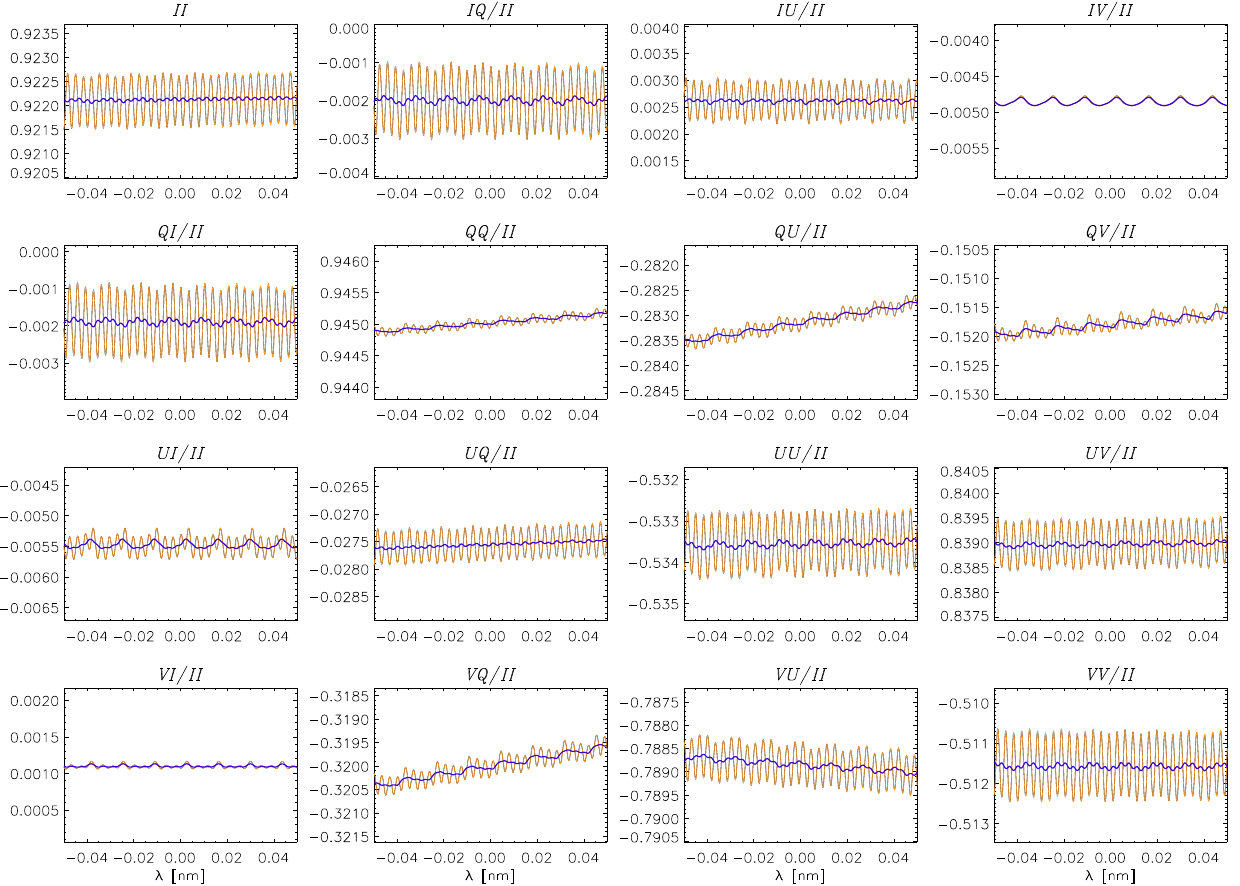}
    \caption{Same PCM model as in Fig.~\ref{fig:Polstar PCM}, but comparing the fully resolved fringes produced in the f/13 configuration (orange curves) with those resulting from the additional smearing by a spectral PSF corresponding to a sampling resolution of 40000 (blue curves).}
    \label{fig:Polstar PCM 2}
\end{figure}

\begin{figure}[t!]
    \centering
\includegraphics[width=\linewidth]{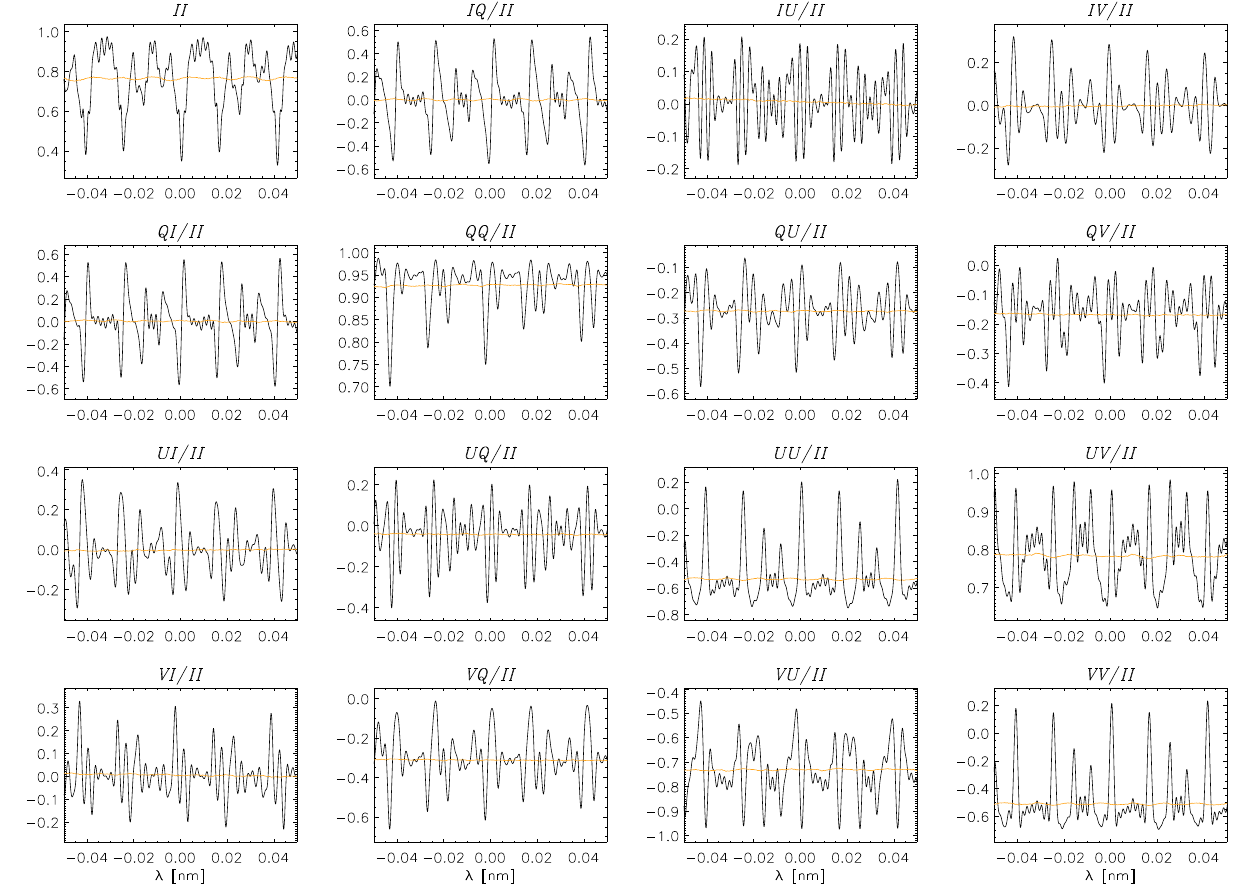}
    \caption{Same as Fig.~\ref{fig:Polstar PCM}, but for an air-gapped assembly of the same modulator, adopting 0.5\,mm spacings between contiguous waveplates.}
    \label{fig:Polstar airgapped}
\end{figure}

Finally, in applications that can rely on low to medium spectral resolution, it is often possible to adjust the overall thickness of an optical compound, possibly including the use of ``passive'' (i.e., non birefringent) elements such as glass windows, in order to decrease the fringe period well below the spectral resolution limit, while at the same time still maintaining relatively low absorptive losses. With this strategy, effective suppression of polarization fringes by spectral smearing can often be attained.

The examples presented in the next section are meant to illustrate these various trades. A practical note about these modeling examples is necessary, which concerns the specification of the polar grid to adopt for optimally sampling the fringe pattern across the beam. Spatial-fringe patterns such as those of Figs.~\ref{fig:spatial_fringes}--\ref{fig:absorb} demonstrate that polarization fringes may become rapidly variable along the angle-of-incidence $\phi$ of the ray within the beam, whereas the $\beta$-azimuth variations around the surface normal typically have a much slower frequency, usually with periods that are multiple of $\pi/2$, as it can be expected for homogeneous A-cut and C-cut birefringent crystals. This suggests that the sampling of the $\beta$-domain can typically be significantly coarser than the sampling of the $\phi$-domain, in order to properly capture the fringe variations along that coordinate, thus allowing for the optimization of the computational effort. For the examples illustrated in the next section, fringe averages over the beam extent appeared to be stable for $\delta\phi\lesssim 0.025^\circ$ and $\delta\beta\lesssim 5^\circ$. Obviously, in the case of less symmetric patterns, such as those that are produced in the presence of an overall tilt of the beam entering the optic, may need significantly finer $\beta$ samplings, and so the choice of the sampling grid must be assessed case by case.

\begin{table}[b!]
    \centering
    \begin{tabular}{l|r|r|r|r}
       \vphantom{$\Bigl.$} 
       & WP \#1 & WP \#2 & WP \#3 & WP \#4 \\
       \hline
       \vphantom{$\Bigl.$}
       Thick. [$\mu$m]  &  403.50 & 400.00 & 410.50 & 400.00 \\
       \hline
       \vphantom{$\Bigl.$}
       Orient. [$^\circ$] & 0.00 & 90.00 & 58.73 & 148.73 \\
    \hline
    \end{tabular}
    \caption{Optical recipe of a PCM optimized between 125 and 285\,nm, employing two MgF$_2$ compound zeroth-order retarders with a bias thickness of 0.4\,mm.}
    \label{tab:PCM}
\end{table}

\subsection{A PCM for the far-UV}

This example is based on the science requirements of a proposed NASA SMEX mission (Polstar \cite{Dr25}), which employs medium-resolution ($R\sim 20000$) spectro-polarimetry to diagnose the magnetism and morphology of massive stars and their environments based on the polarization of prominent spectral lines in their far-UV spectra. The mission operates polarimetrically between 125 and 285\,nm, and achieves nearly optimal polarization modulation efficiency through a modulator design consisting of two compound MgF$_2$ retarders of proper thicknesses. In particular, the model we consider here consists of four MgF$_2$ waveplates, with a minimum plate thickness (\emph{bias thickness}) of 0.4\,mm, and exactly polished to achieve the required retardances. The recipe is given in Table~\ref{tab:PCM}.

\begin{figure}[t!]
    \centering
\includegraphics[width=\linewidth]{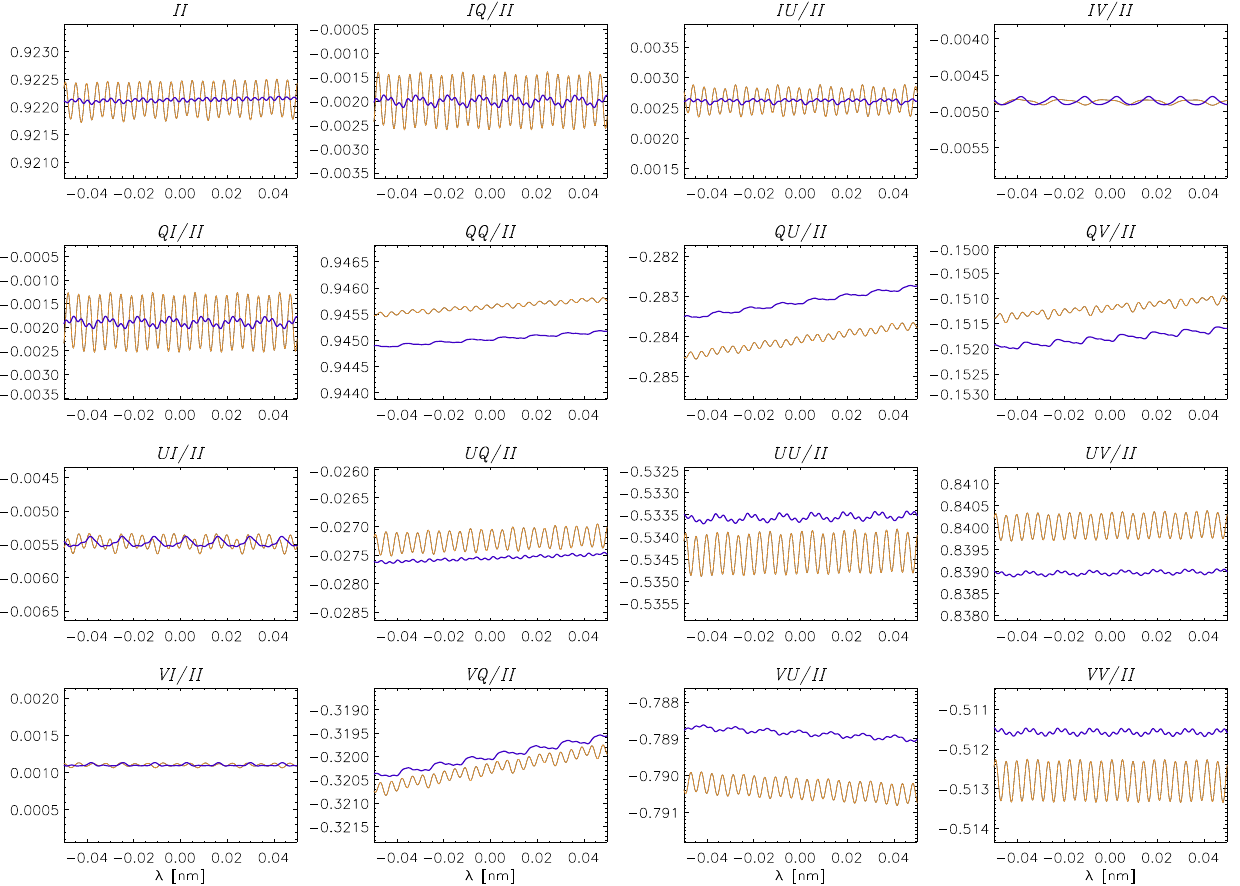}
    \caption{Same modeling of interference fringes as in Fig.~\ref{fig:Polstar PCM 2}, assuming a f/13 beam configuration and a spectral sampling resolution of 40000, using the same PCM model in Table~\ref{tab:PCM} (blue curves; identical solution as the blue curves in Fig.~\ref{fig:Polstar PCM 2}), and after reducing the bias thickness of the four waveplates to 0.3\,mm (orange curves).}
    \label{fig:Polstar PCM 3}
\end{figure}

Figures~\ref{fig:Polstar PCM} and \ref{fig:Polstar PCM 2} illustrate the effects of the first two trades for each of the Mueller matrix elements, showing the fringe patterns that are produced when: 1) the modulator is illuminated by a collimated beam in normal incidence (black curves in Fig.~\ref{fig:Polstar PCM}); 2) the incident beam is converging at f/13 (cone half-aperture of 2.2 deg; orange curves in both Figs.~\ref{fig:Polstar PCM} and \ref{fig:Polstar PCM 2}), in which case the modulator performs a spatial averaging of the fringe patterns corresponding to different ray inclinations through the optic; and 3) after the additional spectral smearing of the fringe pattern (blue curves in Fig.~\ref{fig:Polstar PCM 2}), assuming a spectrograph's PSF corresponding to a sampling resolution of 40000 (a FWHM of about 3.65\,pm). This example clearly shows the cumulative advantage of the spatial and spectral smearings of polarization fringes, when the specific characteristics and performance requirements of an instrument allow us to inform the design of its polarization modulator. 
A similar comparison is shown in Fig.~\ref{fig:Polstar airgapped}, where the modeling conditions are the same as for Fig.~\ref{fig:Polstar PCM}, but the modulator design uses 0.5\,mm thick air gaps to separate the birefringent elements. While the Fresnel reflections at the gap interfaces greatly enhance polarization fringes in a collimated beam, use of a converging/diverging beam through the modulator again significantly depresses the fringe amplitudes. 

In both examples, it is evident how most of the fringe suppression is produced by the spatial smearing of the fringes due to the integration of the signal over the angular extent of the f/13 beam, whereas a much more modest gain in fringe suppression is produced by the additional spectral smearing of the signal because or the  limited resolution of the instrument. In particular, the effect of spectral smearing of the fringes is practically negligible in our example of the air-gapped PCM.

    \label{fig:placeholder}

%

\subsection{Fringe mitigation by material thickness adjustments}

\begin{figure}[t!]
    \centering
\includegraphics[width=\linewidth]{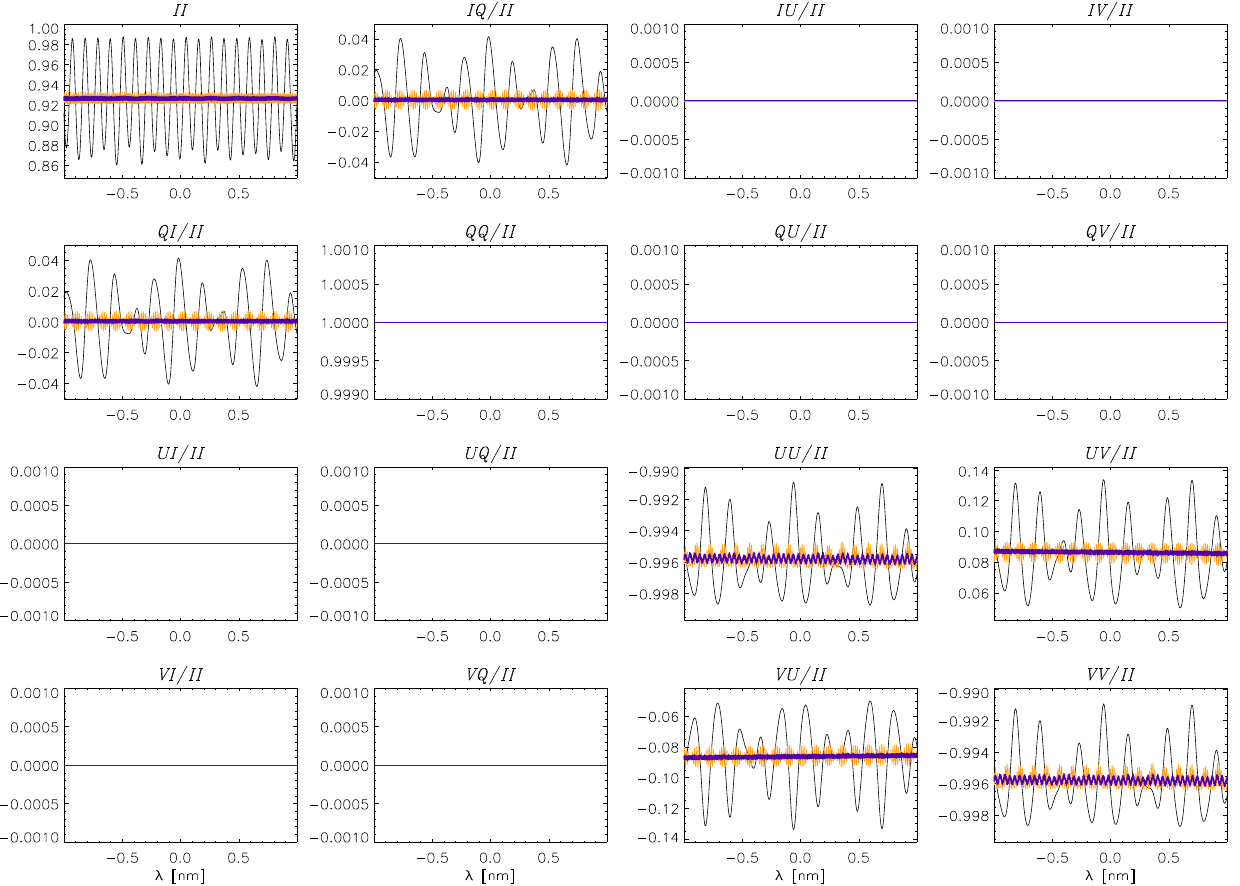}
    \caption{Example of fringe suppression by spectral smearing for the half-wave achromatic retarder of Fig.~\ref{fig:spatial_fringes_achro} in normal incidence around 500\,nm. The black curves show the case of the bare retarder, whereas the colored curves show the case of the retarder sandwiched between index-matching glass windows with 5\,mm (orange) and 15\,mm (blue) thickness. Here we assumed a spectrograph PSF corresponding to a sampling resolution of 20000.}
    \label{fig:Clarke_window}
\end{figure}

When allowed by the spectral requirements of the instrument, the choice of bias thickness of the birefringent elements, compatibly with their fabricability, and of the thickness of the possible gaps between elements (whether these are filled with index matching materials or not), can effectively be used to control the impact of intensity and polarization fringes in the instrument. Similarly, one can often utilize glass windows of adequate thickness to achieve the same results.
The final two examples we present demonstrate the trades of choosing adequate bias thicknesses of birefringent elements and the utilization of windows made of isotropic materials, for the purpose of decreasing the dominant fringe period  well below the spectral resolution of the instrument. 

For the first example, we consider again the FUV PCM modeled in the previous section, and evaluate again the amount of fringe suppression for a f/13 beam of radiation spectrally sampled with an instrument resolution of 40000---similarly to what was shown in Fig.~\ref{fig:Polstar PCM 2}---after reducing the bias thickness of the PCM plates from 0.4\,mm down to 0.3\,mm (cf.~Table~\ref{tab:PCM}). The blue curves in Figs.~\ref{fig:Polstar PCM 2} and \ref{fig:Polstar PCM 3} correspond to the same solution, whereas the orange curves in Fig.~\ref{fig:Polstar PCM 3} now correspond to the case of reduced bias thickness. This comparison shows that, for the adopted shape of the incoming beam and the specified spectral resolution of the instrument, the adoption of thinner plates for the same PCM model leads to a significantly worse suppression of the fringe pattern over the wavelength interval of this example.

For the second example, we consider again the  SiO$_2$-MgF$_2$ achromatic half-wave retarder used for Fig.~\ref{fig:spatial_fringes_achro}, illuminated by a collimated beam in normal incidence around 500\,nm. The three different curves in the Mueller matrix plots of Fig.~\ref{fig:Clarke_window} show the cases of the original achromat model (black curves), and for the same achromat sandwiched between two index-matching glass windows (nominally, amorphous SiO$_2$ and MgF$_2$ crystals) with a thickness of 5\,mm (orange curves) and 15\,mm (blue curves). The sampling resolution of the plot is 20000, for an effective spectral resolution of about 10000.
We note the significant suppression of polarization fringes that can be attained by utilizing passive windows of adequate thickness in applications that do not require a very high spectral resolution.

\subsection{Other considerations about fringe mitigation}

The techniques discussed above for suppressing intensity and polarization fringes in spectro-polarimetric data will generally have an impact on the optical performance of the instrument they are implemented in. This is after all an inevitable aspect of the trades we have discussed. However, it is critical to realize that, while such a performance impact can always be addressed---with the investment of sufficient resources---in the design and engineering phases of instrument development, the impact of fringes on the scientific viability of spectro-polarimetric data is often destructive and irrecoverable. Despite the fact that several signal-filtering techniques have been proposed over the years to curb \emph{ex post facto} the detriment from polarization fringes on science data (see, e.g., \cite{CL19} and references therein), such techniques can only be applied in particularly benign situations, e.g.: the fringes must have a sufficiently low amplitude compared to the science signals; the power spectra of the science data and the fringes must have largely non-overlapping frequency ranges; the phases of the fringe pattern must be such that no amplitude peak/trough of the fringes happens to coincide with the spectral frequencies of the science signals that are critical for an accurate interpretation of the acquired data. These conditions, unfortunately, are the exception rather than the rule, and often the fact that fringe patterns may satisfy such benign conditions is already the result of some mitigation approach to the issue.

Below we enumerate typical examples of performance concerns that must be addressed in practical applications of fringe-mitigation techniques. These have already been analyzed and discussed in greater detail elsewhere, so we limit ourselves to referencing the relevant literature.

\begin{enumerate}
\item The placing of a retarder in a converging or diverging beam introduces concerns about the retarder's spatial uniformity and FOV dependence of its calibration \cite{JATIS8}. This can be mitigated in the design phase of the retarder, as well as through the specification of adequate fabrication tolerances, the adoption of accurate metrology equipment to measure retarder variation, and the use of appropriate data reduction algorithms \cite{JATIS4,JATIS9,SolPhysDMH}. 

\item The suppression of fringes by averaging over the angular span of a converging or diverging beam will in general be accompanied by some degree of depolarization (see Sect.~5 of \cite{Ha23}, and Ch.~6 of \cite{ChipmanBook}). Therefore, depolarization must be modeled during system design, to ensure amplitudes remain below the allocated error budget. Common error budgets in solar telescopes adopt depolarization limits near 1\% (e.g. DKIST \cite{SolPhysDMH}, Hinode \cite{Ichimoto08}). It is also important to remark that depolarization can easily be measured in the lab to validate models (c.f.~\cite{Ha23}, App.~B), and if necessary it can be accounted for in calibration algorithms by including appropriate variables in the system model \cite{LuChipman1996, Chipman05, ossikovski2009, Noble2012, Ha23}.

\item The use of glass windows of sufficient thickness to manipulate the fringe period magnifies the risks associated with beam wobble during the rotation of the optic through the modulation cycle. This can be mitigated by imposing tight rotary stage fabrication tolerances and using high-accuracy optical alignment equipment such as laser trackers \cite{JATIS11}.  
\end{enumerate}

\section{Conclusions}

We used known results on the propagation of waves through anisotropic materials \cite{He91,Di63,Ye82,Cl04c} as the basis for the derivation of an approximate Jones formalism that allows us to rapidly model polarization fringes in a stack of birefringent materials, including the case of illumination by a finite f/\# beam, with sufficient accuracy for optical design trade studies. 
Our approach is capable to reproduce results presented in previous literature on the modeling of \emph{spectral} fringes \cite{Cl04a,Cl04b,Cl04c}, as well as to model the formation of \emph{spatial} fringes in optics illuminated by a non-collimated beam \cite{Ha23}. Examples of both types of modeling have been presented in this paper. 

The modeling framework developed in this work  may not allow us to reproduce \emph{all} numerical details of more rigorous treatments such as Berreman calculus \cite{Be72,Ye82,MC15,ChipmanBook}, but nonetheless it still fulfills the purpose of effectively predicting both pattern and magnitude of the polarization fringes produced in the presence of plane-parallel optics inside an instrument. The accuracy of our approximate treatment is completely adequate to quantitatively assess the impact of such fringes on the polarimetric performance of the instrument, based on realistic instrument design parameters, such as spectral resolution and f/\# \cite{Ha23}, and how those design parameters need to be modified in order to reduce the impact of fringes. It therefore provides a convenient and sufficiently reliable tool for the validation of optical-design strategies of fringe mitigation for different polarimetric applications and instrument concepts. To demonstrate this, we modeled various types of birefringent optics, and showed in detail how fringes can be manipulated and effectively suppressed in the observed data by making informed design choices. In fact, this type of modeling has already been used to guide the optical design of major nationally funded observing facilities, such as the DKIST \cite{JATIS2,JATIS3,JATIS8}, in order to quantify and mitigate the impact of these polarimetric artifacts on the science deliverables of the project.

As a final note, we point out how this type of modeling also allows us to simulate the manifestation of polarization fringes directly in the science data. While we have not provided a practical example of this, we briefly describe the concept here. In order to realistically simulate the polarimetric performance of an instrument, one must model its polarization \emph{modulation} matrix \cite{To10}, properly averaged over the angular range of the incoming optical beam of the instrument and at the proper spectral sampling resolution (see, e.g., Fig.~\ref{fig:Polstar PCM 2}). When such a modulation matrix is adopted for the data simulation pipeline of the instrument, which maps the incoming Stokes vector from a polarized target to the modulated signals measured by the detector, the simulated signals will be affected by the presence of spatial and spectral fringes. On the other hand, the data reduction for a polarimetric instrument involves the use of a reference \emph{demodulation} matrix that decodes the modulated signals into the Stokes profiles of the target. Such a demodulation matrix must be determined via polarization calibration of the instrument under environmental conditions that ideally should be identical to those of the observations. Because such an ideal occurrence is never realized in practice, and fringes are highly sensitive to even small variations of the environmental conditions, the product of the reference demodulation matrix with the actual modulation matrix generally does not correspond to an identity, the residuals being responsible for the introuction of polarization errors such as cross-talk \cite{Ca25} and FOV-dependent fringes. In other words, the ``blind'' demodulation of signals produced by a ``fringed'' modulation matrix will naturally cause the appearance of fringe patterns that are superimposed on the true polarization signals from the source. This type of modeling can be used to test modulator designs specifically aimed at mitigating the effects of polarization fringes on Stokes measurements, and also to create synthetic observations with fringes that can be used to test the applicability and effectiveness of post-processing de-fringing methods \cite{CL19} or the resilience of Stokes inversion codes to the presence of fringes in the data. 


We expect that such in-depth design analyses will become increasingly relevant for the development of future instrumentation, and envision that the use of modeling tools such as the one presented in this work will become commonplace in the design phase of any polarimetric instrument.

\bigskip \noindent
\textbf{Acknowledgments and disclosures.}
This material is based upon work supported by the NSF National Center for Atmospheric Research, which is a major facility
sponsored by the National Science Foundation under Cooperative Agreement No.~1852977. DMH acknowledges support by the NSF Daniel K. Inouye Solar Telescope (DKIST) project. 
The authors declare no conflicts of interest.

\bigskip \noindent
\textbf{Data availability.}
Data underlying the results presented in this paper were exclusively modeled. They are not publicly available at this time, but they may be obtained from the authors upon reasonable request. The birefringence data of SiO$_2$ and MgF$_2$ used for the models in this work rely on the measurements by S.~Sueoka and D.~Elmore for the DKIST project \cite{Su16}.

\appendix

\numberwithin{equation}{section}
\setcounter{equation}{0}

\section{Alternative derivation of the transfer law}
\label{sec:appA}

Following the formalism presented in Sects.~\ref{sec:formalism} and \ref{sec:rotations}, we indicate with $(\bm{\hat J}^+_\alpha,\bm{\hat J}^-_\alpha)^T_{m}$
the Jones 4-vector in the $m^\textrm{th}$ medium, expressed in the frame of reference of the principal axes of the same medium, and evaluated at the interface with the $(m+1)^\textrm{th}$ medium. We have (cf.~Eq.~(\ref{eq:phase-jones}))
\begin{displaymath}
\begin{pmatrix}
\bm{\hat J}^+_\alpha \\
\bm{\hat J}^-_\alpha
\end{pmatrix}_{m}
=
\mathbf{\Lambda}_{m}
\begin{pmatrix}
\bm{J}^+_\alpha \\
\bm{J}^-_\alpha
\end{pmatrix}_{m}\;,
\end{displaymath}
where (cf.~Eq.~(\ref{eq:rotated-jones}))
\begin{displaymath}
\begin{pmatrix}
\bm{J}^+_\alpha \\
\bm{J}^-_\alpha
\end{pmatrix}_{m}
=\mathbf{Q}(\alpha_{m})
\begin{pmatrix}
\bm{J}^+ \\
\bm{J}^-
\end{pmatrix}_{m}\;,
\end{displaymath}
and similarly for $(\bm{\hat J}^+_\alpha,\bm{\hat J}^-_\alpha)^T_{m}$. Combining these two equations, we find at once
\begin{equation} \label{eq:alt1}
\begin{pmatrix}
\bm{\hat J}^+ \\
\bm{\hat J}^-
\end{pmatrix}_{m}
=
\mathbf{Q}(-\alpha_{m})
\mathbf{\Lambda}_{m}
\mathbf{Q}(\alpha_{m})\,
\begin{pmatrix}
\bm{J}^+ \\
\bm{J}^-
\end{pmatrix}_{m}\;.
\end{equation}
At the same time, the transfer law across the interface between the $(m-1)^\textrm{th}$ and $m^\textrm{th}$ materials, expressed in the 
reference frame of the principal axes of the $m^\textrm{th}$ medium, implies that (cf.~Eq.~(\ref{eq:temp}))
\begin{displaymath}
\mathbf{Q}(\alpha_{m})
\begin{pmatrix}
\bm{\hat J}^+ \\
\bm{\hat J}^-
\end{pmatrix}_{m-1}
=
\mathbf{O}_{m-1,m}\,
\mathbf{Q}(\alpha_{m})
\begin{pmatrix}
\bm{J}^+ \\
\bm{J}^-
\end{pmatrix}_m
\;,
\end{displaymath}
which, together with Eq.~(\ref{eq:alt1}), gives
\begin{equation} \label{eq:alt2}
\begin{pmatrix}
\bm{J}^+ \\
\bm{J}^-
\end{pmatrix}_{m-1}
=
\bigl[\mathbf{Q}(-\alpha_{m-1})
\mathbf{\Lambda}_{m-1}^\ast
\mathbf{Q}(\alpha_{m-1})\bigr]
\bigl[\mathbf{Q}(-\alpha_m)
\mathbf{O}_{m-1,m}
\mathbf{Q}(\alpha_m)\bigr]
\begin{pmatrix}
\bm{J}^+ \\
\bm{J}^-
\end{pmatrix}_m\;,
\end{equation}
to be compared with the result of Eq.~(\ref{eq:Ye82}). 

For a stack of $N$ materials, Eq.~(\ref{eq:stacksolve.1}) still applies, with $\mathscr{M}_m$ now being given by
\begin{equation} \label{eq:altM}
\mathscr{M}_m
\equiv
\bigl[\mathbf{Q}(-\alpha_{m-1})
\mathbf{\Lambda}_{m-1}^\ast
\mathbf{Q}(\alpha_{m-1})\bigr]
\bigl[\mathbf{Q}(-\alpha_m)
\mathbf{O}_{m-1,m}
\mathbf{Q}(\alpha_m)\bigr]\;,
\end{equation}
instead of Eq.~(\ref{eq:stack}).

The numerical implementation of this alternative formulation gives results identical to those presented in this work. In fact, it is easy to show that Eq.~(\ref{eq:Ye82}) transforms into Eq.~(\ref{eq:alt2}) when the relation (\ref{eq:alt1}) is used in both sides of the equation.


\section{Numerical implementation} 
\label{app:numerical}

The development of the formalism presented in this paper relies on two critical approximations: 1) the physical condition of small birefringence for the materials to be modeled; and 2) the assumption that the transfer matrix $\mathbf{O}_{m-1,m}$ can be expressed through the usual expressions of the Fresnel coefficients for the transmission and reflection of the $p$ and $s$ components of the radiation field (e.g., Eqs.~4(19--22) of \cite{He91}), where the (generally complex) refractive indexes of the preceding material $(m-1)$ are referenced to the principal axes of the current material $m$. 
We are not going to address here the approximation of small birefringence (see App.~\ref{app:gen_biref}). 
Instead, we provide a compendium of the formulas taken from \cite{He91} that are relevant for the numerical implementation of the formalism presented in this work, using the notation of this paper, and illustrating the necessary generalizations.

We first want to express the Fresnel coefficients for the reflection and transmission of a plane wave at the interface between two birefringent media $(m-1)$ and $m$ in the reference frame of the principal axes of the $m^\textrm{th}$ medium. These coefficients are specified for two orthogonal states of polarization (cf.~Eq.~(\ref{eq:tranmat4}) and following discussion). For an ``A-cut'' uniaxial crystal, one polarization lies on the $\langle\bm{o},z\rangle$ plane of the $m^\textrm{th}$ medium, whereas the perpendicular polarization is parallel to the $\bm{e}$-axis. 
Following Eqs.~4(19--22) of \cite{He91}, we can write
\begin{eqnarray} \label{eq:Fresnel}
r^{\rm o}_{m-1,m}&=&\frac{\tilde\nu_{m-1}^{\rm o}\cos\phi_m^{\rm o}-\nu_m^{\rm o}\cos\phi_{m-1}^{\rm o}}{\tilde\nu_{m-1}^{\rm o}\cos\phi_m^{\rm o}+\nu_m^{\rm o}\cos\phi_{m-1}^{\rm o}}\;, \\
t^{\rm o}_{m-1,m}&=&\frac{2\,\tilde\nu_{m-1}^{\rm o}\cos\phi_{m-1}^{\rm o}}{\tilde\nu_{m-1}^{\rm o}\cos\phi_m^{\rm o}+\nu_m^{\rm o}\cos\phi_{m-1}^{\rm o}}\;, \\ 
\noalign{\allowbreak}
r^{\rm e}_{m-1,m}&=&\frac{\tilde\nu_{m-1}^{\rm e}\cos\phi_{m-1}^{\rm e}-\nu_m^{\rm e}\cos\phi_m^{\rm e}}{\tilde\nu_{m-1}^{\rm e}\cos\phi_{m-1}^{\rm e}+\nu_m^{\rm e}\cos\phi_m^{\rm e}}\;, \\
\label{eq:Fresnel.last}
t^{\rm e}_{m-1,m}&=&\frac{2\,\tilde\nu_{m-1}^{\rm e}\cos\phi_{m-1}^{\rm e}}{\tilde\nu_{m-1}^{\rm e}\cos\phi_{m-1}^{\rm e}+\nu_m^{\rm e}\cos\phi_m^{\rm e}}\;,
\end{eqnarray}
%
%
where we indicated with $\tilde\nu_{m-1}^{\rm o,e}$ the (complex) indexes of refraction of the $(m-1)^\textrm{th}$ birefringent medium that have been ``resolved'' to the principal-axes frame of the $m^\textrm{th}$ medium, according to the prescription that we will be presenting below. The propagation angles $\phi^{\rm o,e}$ satisfy the (complex) Snell's law (\ref{eq:snell}). As we pointed out in Sect.~\ref{sec:rotations}, these propagation angles do not exactly define the real direction of the propagating ray in the case of a conducting medium, which is characterized by a complex index of refraction. If we indicate with $\hat\nu=\hat n-\imag \hat k$ the complex index of refraction actually experienced by the ray propagating through the conducting medium at an angle $\hat\phi$ from the normal to the interface, this index is related to the $(n,k)$ pair specifying the material's complex index via the following three non-linear relations (cf.~Eqs.~4(4-6) of \cite{He91}; also, Eqs.~15(101,102) and 15(98) of \cite{Di63})
\begin{eqnarray}
    {\hat n}^2-{\hat k}^2 &=& n^2-k^2\;, \\
    \hat n\,\hat k\cos\hat\phi &=& n\,k\;, \\
    \hat n\sin\hat\phi &=& \hat n_0\sin\hat\phi_0\;,
\end{eqnarray}
where the 0-subscripted quantities correspond to the  preceding medium. If so desired, this set of relations can simply be solved with an iterative scheme, which is found to converge very rapidly for the typical values of the extinction coefficient $k$ of conducting materials.

In turn, the refractive indexes experienced by an incoming (polarized) ray are generally dependent on the angle of propagation $\phi$ through the medium, which affects the values that must be adopted for the $\nu^{\rm o,e}$ indexes in Eqs.~(\ref{eq:Fresnel}--\ref{eq:Fresnel.last}),
\footnote{This is commonly described as a condition of non-validity of Snell's law for the ray that experiences a direction-dependent refractive index.} 
the derivation of which involves the solution of a bi-quadratic equation for the index ellipsoid of the material (cf.~Eq.~(2.61) of \cite{MC15}). For an ``A-cut'' uniaxial crystal rotated by $\alpha$ with respect to the $x$-axis, we find
\begin{equation} \label{eq:ne_update}
\mathrm{Re}({\nu}^{\rm o})=n_{\rm o}\;, \qquad
\mathrm{Re}({\nu}^{\rm e})=
\frac{n_{\rm e} n_{\rm o}}{\sqrt{n_{\rm o}^2 \cos^2\phi^{\rm e}+\bigl(n_{\rm o}^2 \cos^2\alpha+n_{\rm e}^2\sin^2\alpha\bigr)\sin^2\phi^{\rm e}}}\;.
\end{equation}
This generalization of $\nu^{\rm e}$ in the case of non-normal incidence represents a first-order relaxation of the approximation of small birefringence, which is readily implementable in our formalism, allowing us to model rotational asymmetries of the spatial fringes produced in anisotropic materials illuminated by a finite f/\# beam. Examples of these asymmetries are clearly seen in the $UV$ and $VU$ elements of the Mueller maps of Figs.~\ref{fig:spatial_fringes} and \ref{fig:spatial_fringes_achro}. When the small-birefringence approximation $n_{\rm e}\approx n_{\rm o}$ is made in the denominator of Eq.~(\ref{eq:ne_update}), such a relation gives $\mathrm{Re}(\nu^{\rm e})\approx n_e$. We note that this becomes an exact condition when the birefringent material is aligned with the PoI ($\alpha=0$) or in normal incidence ($\phi^{\rm o,e}=0$).

For a ``C-cut'' uniaxial crystal, the $\bm{e}$-axis is normal to the interface, so the material is rotationally invariant around the optic axis. The two solutions for the refractive index seen by an incoming ray can then be distinguished simply based on the orientation of the PoI, and they become
\begin{equation} \label{eq:Zcut}
\mathrm{Re}(\nu^p)=
\frac{n_{\rm o} n_{\rm e}}{\sqrt{n_{\rm e}^2 \cos^2\phi^p+n_{\rm o}^2 \sin^2\phi^p}}\;, \qquad
\mathrm{Re}(\nu^s)=n_{\rm o}\;.
\end{equation}
%
Similarly, Eqs.~(\ref{eq:Fresnel}--\ref{eq:Fresnel.last}) remain formally identical with the substitution ${\rm (o,e)}\to(p,s)$.

Next we provide the recipe to ``resolve'' the refractive indexes in the Fresnel formulas (\ref{eq:Fresnel}--\ref{eq:Fresnel.last}) that apply at the interface between differently oriented birefringent materials. Let $n_{\rm o,e}$ be the (real, positive) 
refractive indexes of the $(m-1)^\textrm{th}$ birefringent material expressed in its system of principal axes.
We may then assume the following ``resolution'' rule of such indexes to the principal axes of the $m^\textrm{th}$ material, rotated by an angle $\Delta\alpha=(\alpha_m-\alpha_{m-1})$ around the surface normal with respect to the $(m-1)^\textrm{th}$ material (see Fig.~\ref{fig:perp_geom}):
%
%
\begin{equation} 
\label{eq:resolution}
(\tilde n_{\rm o,e})^2 = (n_{\rm o,e} \cos\Delta\alpha)^2+ (n_{\rm e,o} \sin\Delta\alpha)^2\;.
\end{equation}
A justification for this rule lies in the form of the permittivity tensor projected on the $\langle x,y\rangle$ plane, and expressed in the reference frame of the principal axes of the $(m-1)^\textrm{th}$ material, which is simply a 2$\times$2 diagonal matrix with elements $n_{\rm o}^2$ and $n_{\rm e}^2$. Upon transformation to the principal-axes frame of the  material $m^\textrm{th}$, the new tensor is nearly diagonal under the approximation of small birefringence, the diagonal elements being the squares of new indexes ${\tilde n}_{\rm o,e}$ that are obtained through a rule identical to Eq.~(\ref{eq:resolution}),
while the off-diagonal elements are simply proportional to the (small) birefringence $(n_{\rm e}-n_{\rm o})$.

An alternative recipe can be provided based on the idea of ``resolution'' of the group velocities rather than the refractive indexes:
%
%
\begin{equation} 
\label{eq:resalt}
\biggl(\frac{1}{\tilde n_{\rm o,e}}\biggr)^2 = \biggl(\frac{\cos\Delta\alpha}{n_{\rm o,e}}\biggr)^2+
\biggl(\frac{\sin\Delta\alpha}{n_{\rm e,o}}\biggr)^2\;.
\end{equation}
For all the examples presented in this work, we found that the differences in the results by adopting Eq.~(\ref{eq:resalt}) instead of Eq.~(\ref{eq:resolution})
are largely insignificant when compared to other approximations implied by our formalism.

\begin{figure}[t!]
    \centering 
\includegraphics[width=\textwidth]{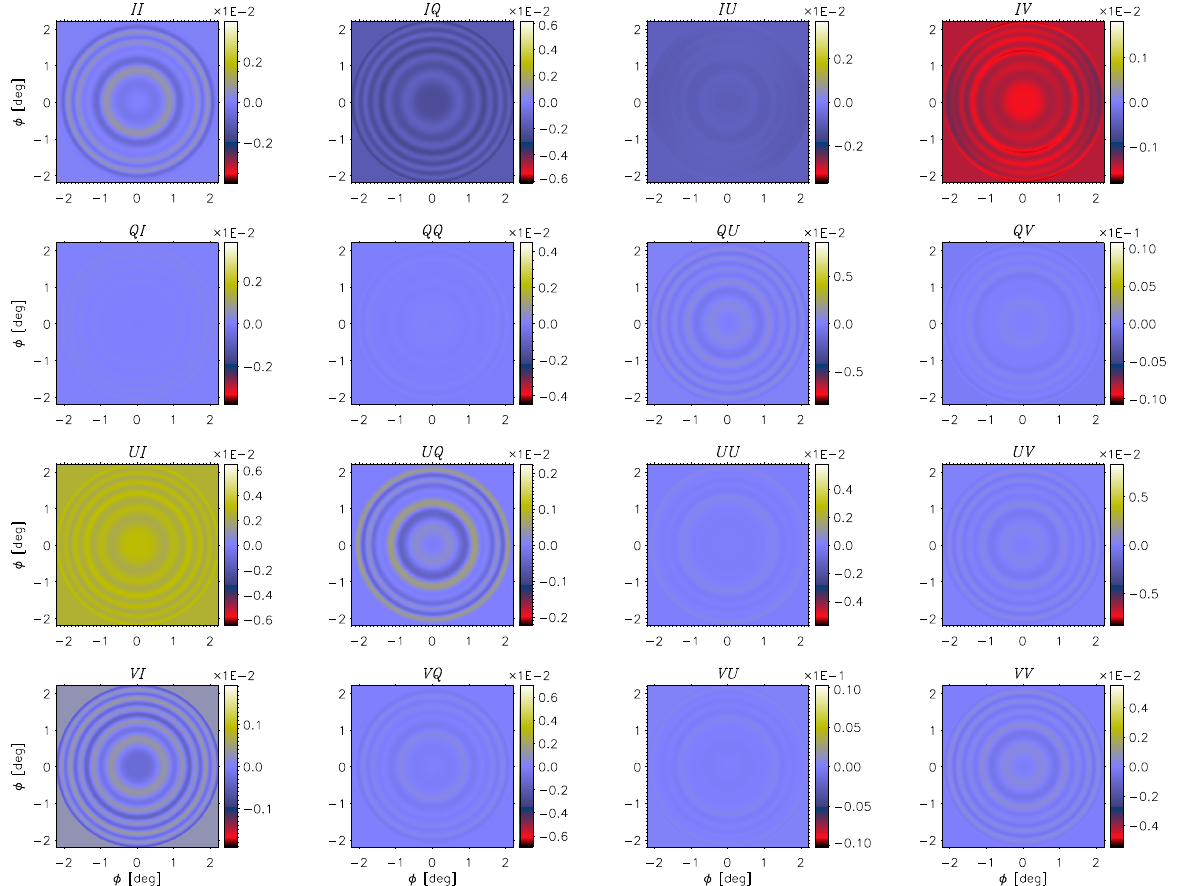}
    \caption{
    Difference maps between two calculations of the Mueller matrix of the PCM of Fig.~\ref{fig:check}, illuminated by a f/13 beam of monochromatic light at 500\,nm. The first calculation assumes the exit vacuum is ``aligned'' with the PoI, and the second one assumes that the exit vacuum is ``aligned'' with the last birefringent optic in the stack. The largest deviations are at most a few percent of the fringe amplitudes in the Mueller matrix, with more evident offsets in the diattenuation and polarizance vectors.}
    \label{fig:align_error}
\end{figure}

If the absorptivity of the birefringent material can be assumed to be isotropic, 
Eq.~(\ref{eq:resolution}) (or the alternative Eq.~(\ref{eq:resalt}))
is the only ``resolution'' rule for the refractive indexes that must be considered; otherwise, analogous expressions must be separately adopted for the distinct $\tilde k_{\rm o,e}$ imaginary coefficients.

The above recipes allow us to calculate the transfer matrix $\mathbf{O}_{m-1,m}$ of Eq.~(\ref{eq:stack}), and they lead to results that are in very good agreement with those presented in previous literature (e.g., compare the examples of Fig.~\ref{fig:examples} with the various waveplate models considered by \cite{Cl04a,Cl04b}), or computed using exact formulations such as \cite{Be72}.


Another practical issue of our treatment concerns the closure of the recursive relation Eq.~(\ref{eq:stack}) with the vacuum as the $(N+1)^\textrm{th}$ material. Let us consider the simplest case of one birefringent optic immersed in the vacuum in normal incidence, so Eq.~(\ref{eq:stacksolve.1}) explicitly reads
\begin{eqnarray}
\begin{pmatrix}
\bm{J}^+ \\
\bm{J}^-
\end{pmatrix}_{0}
&=&
\left[ \mathbf{Q}(-\alpha_1) \mathbf{O}_{0,1}\mathbf{\Lambda}_1^\ast \mathbf{Q}(\alpha_1)\right]
\left[ \mathbf{Q}(-\alpha_2) \mathbf{O}_{1,2}\mathbf{\Lambda}_2^\ast \mathbf{Q}(\alpha_2) \right]\mathbf{\Lambda}_2
\begin{pmatrix}
\bm{J}^+ \\
\bm{0}
\end{pmatrix}_{2} \nonumber \\
&=&\left[ \mathbf{Q}(-\alpha_1) \mathbf{O}_{0,1}\mathbf{\Lambda}_1^\ast \mathbf{Q}(\alpha_1)\right]
\left[ \mathbf{Q}(-\alpha_2) \mathbf{O}_{1,2} \mathbf{Q}(\alpha_2) \right] 
\begin{pmatrix}
\bm{J}^+ \\
\bm{0}
\end{pmatrix}_{2}\;,
\end{eqnarray}
where for the second equivalence we used the fact that the vacuum is isotropic, and thus $\mathbf{\Lambda}_2$ is simply proportional to the unit matrix and commutes with any matrix. Once again, we point out that the exit Jones 4-vector $(\bm{J}^+,\bm{0})_2^T$ is already properly expressed in the PoI reference frame (see Fig.~\ref{fig:perp_geom} and Sect.~\ref{sec:rotations}).

Because $\mathbf{Q}(\alpha_2)$ generally does not commute with $\mathbf{O}_{1,2}$, unless the optic is isotropic \emph{and} illuminated in normal incidence,\footnote{Commutativity in the case of isotropic materials is numerically verified as long as one can assume $\cos\phi\sim 1$, which holds with an error of 1\% (0.1\%) for light beams as fast as $\sim$ f/3.5 (f/11.2).} the choice of $\alpha_2$ along which the vacuum is ``oriented'' is not arbitrary in our formalism, despite the vacuum being isotropic! 
Two among all possible orientation ``choices'' stand out: 1) to ``orient'' the exit vacuum so it is aligned with the PoI frame, i.e., $\alpha_2=0$; and 2) to set $\alpha_2=\alpha_1$, in order to eliminate a formally redundant rotation from the last optic's principal-axes frame when exiting into the vacuum.

Figure~\ref{fig:align_error} shows the difference maps of the spatial fringes for the Mueller matrix of the PCM of Fig.~\ref{fig:check} evaluated at 500\,nm, using the first orientation choice, $\alpha_{N+1}=0$, and subtracting 
the one obtained by adopting the second orientation choice, $\alpha_{N+1}=\alpha_N$.
For each element of this difference matrix, the bounding interval of the plot is set to 10\% of the  amplitude min-max range of the corresponding element in the Mueller matrix, and it is centered around zero to emphasize offsets; for example, if a given element of the Mueller matrix has an amplitude range of 0.1, 
the range of the corresponding element in the difference matrix is set to 
$(-0.005,+0.005)$,
for a total span of 0.01.
As the deviations of the fringe patterns between the two configurations appear to remain well below the saturation level corresponding to the bounding intervals, we can conclude that these deviations are at most of a few percent. As seen from the figure, the largest ones appear to be offsets that impact the diattenuation and polarizance vectors of the Mueller matrix. 


\emph{We trace these discrepancies back to our approximate treatment of the formation of spectral and spatial fringes in a stack of birefringent optics, and interpret them as a direct quantification of the impact of the approximations involved.} In our limited tests, the condition $\alpha_{N+1}=\alpha_N$ appeared to lead to results in better agreement with the Berreman calculus, and to satisfy more accurately the condition of energy conservation (transmittance $\le 100\%$). For the models shown in Fig.~\ref{fig:check}  and in Figs.~\ref{fig:Polstar PCM} through \ref{fig:Polstar airgapped}
we have adopted this orientation choice.
%
When the condition $\alpha_{N+1}=\alpha_N$ is adopted, for consistency, the same rule should also be applied to any other isotropic material possibly present \emph{between} the last birefringent optic and the exit vacuum.\footnote{For an isotropic material between two birefringent optics, we have found numerically that, in the limit of vanishing thickness, the algorithm converges somewhat more accurately to the expected result for that material being absent, if the material is ``aligned'' to the \emph{nearest following birefringent optic}. If so desired, this can be implemented as a standard rule of the algorithm, in which case, by extension, it must also be applied to any isotropic materials \emph{preceding} the first birefringent optic in the stack.}

Finally, when the stack does not include any birefringent optics, the algorithm gives more accurate results when all rotation matrices are taken as the identity (i.e., all isotropic materials are ``aligned'' with the PoI).
%
The absorptive case of Fig.~\ref{fig:absorb} was  calculated using such a setup.

\section{Transformation matrix for arbitrary birefringence} \label{app:gen_biref}

For a strongly birefringent medium, characterized by a significant
difference between the $\nu_{\rm o}$ and $\nu_{\rm e}$ values of the (complex)
indexes of refraction along the ordinary and extraordinary
axes, the polarizations of the o and e rays away from normal incidence can no longer be considered to be approximately orthogonal
to each other (see, e.g., Eqs.~(16,17) of \cite{Ye82}, and discussion thereafter), and therefore the transformation through the matrix $\mathbf{Q}$ in
Eq.~(\ref{eq:stack}) no longer corresponds to a pure rotation of Jones vectors.
In such a case, the rotation matrix $\mathbf{R}$ in the definition 
(\ref{eq:defQ}) of the $\mathbf{Q}$ matrix generalizes to the 
following (cf.~Eqs.~(42--45) of \cite{Ye82}; see also Eq.~(\ref{eq:defR}) and Fig.~\ref{fig:perp_geom}),
\begin{equation} \label{eq:gen_R}
\mathbf{R}=
\begin{pmatrix}
\bm{o\cdot p} &\bm{o\cdot s} \\
\bm{e\cdot p} &\bm{e\cdot s}
\end{pmatrix}\;.
\end{equation}
%
While it is still true that (cf.~Eqs.~(\ref{eq:projected}))
\begin{equation} \label{eq:o-ref}
\bm{o\cdot p}=\cos\psi\;,\qquad
\bm{o\cdot s}=\sin\psi\;,
\end{equation}
we now have instead
\begin{equation} \label{eq:e-ref}
\bm{e\cdot p}=\cos(\psi+\eta)\;,\qquad
\bm{e\cdot s}=\sin(\psi+\eta)\;,
\end{equation}
where $\eta$ is the angle between the $\bm{o}$ and $\bm{e}$ polarizations, which generally is no longer equal to $\pi/2$ for an oblique direction of propagation. As expected, for $\eta\to\pi/2$, Eq.~(\ref{eq:gen_R}) transforms into the rotation operator Eq.~(\ref{eq:defR}), which is the condition implied by the small-birefringence approximation.. 


In order to demonstrate Eqs.~(\ref{eq:e-ref}), we note that, by definition of the angle $\eta$,
\begin{equation}
\bm{o\cdot e}=\cos\eta\;,\qquad
\bm{o\times e}=\sin\eta\,\bm{k}\;,
\end{equation}
where $\bm{k}$ is the propagation unit vector. Using the vector identities
\begin{equation} \label{eq:vecident}
\bm{o}=
(\bm{o\cdot p})\,\bm{p}+(\bm{o\cdot s})\,\bm{s}\;, \qquad
\bm{e}=
(\bm{e\cdot p})\,\bm{p}+(\bm{e\cdot s})\,\bm{s}\;,
\end{equation}
and recalling Eq.~(\ref{eq:o-ref}), we have
\begin{eqnarray*} 
\cos\eta=\bm{o\cdot e}
&=&(\bm{o\cdot p})\,(\bm{e\cdot p})+(\bm{o\cdot s})\,(\bm{e\cdot s}) \\
&=&(\bm{e\cdot p})\cos\psi+(\bm{e\cdot s})\sin\psi\;, \\
\sin\eta=(\bm{o\times e})\cdot\bm{k}
&=&(\bm{o\cdot p})\,(\bm{p\times e})\cdot \bm{k}+(\bm{o\cdot s})\,(\bm{s\times e})\cdot \bm{k}  \\
&=&(\bm{e\cdot s})\cos\psi-(\bm{e\cdot p})\sin\psi\;,
\end{eqnarray*}
where for the last equivalence we simply took the cross product of $\bm{p}$ and $\bm{s}$, respectively, with the expression of $\bm{e}$ in Eq.~(\ref{eq:vecident}), and also noted that $\bm{p\times s}=\bm{k}$. The last two equations form a linear system in the variables $\bm{e\cdot p}$ and $\bm{e\cdot s}$, which is readily solved to give
\begin{eqnarray*}
\bm{e\cdot p}
&=&-\sin\psi\sin\eta+\cos\psi\cos\eta=\cos(\psi+\eta)\;, \\
\bm{e\cdot s}
&=&\cos\psi\sin\eta+\sin\psi\cos\eta=\sin(\psi+\eta)\;,
\end{eqnarray*}
which correspond to Eqs.~(\ref{eq:e-ref}). 

Because $\mathbf{R}$ is no longer an orthogonal transformation, its inverse cannot be obtained simply via transposition, and we find instead that
\begin{equation} \label{eq:gen_Rinv}
\mathbf{R}^{-1}=\frac{1}{\sin\eta}
\begin{pmatrix}
\bm{e\cdot s} &-\bm{o\cdot s} \\
-\bm{e\cdot p} &\bm{o\cdot p}
\end{pmatrix}\;,
\end{equation}
along with Eqs.~(\ref{eq:o-ref}) and (\ref{eq:e-ref}).

\end{document}